\documentclass[twocolumn]{aastex63}

\usepackage{graphicx}	
\usepackage{amsmath}	
\usepackage{amssymb}	
\usepackage{xspace}

\newcommand{\Ha}{H$\alpha$\xspace}
\newcommand{\Hb}{H$\beta$\xspace}
\newcommand{\Hd}{H$\delta$\xspace}
\newcommand{\Msun}{$\rm M_\odot$\xspace}
\newcommand{\kms}{$\rm km \, s^{-1}$\xspace}
\newcommand{\Hunit}{$\rm km \, s^{-1} \,  Mpc^{-1}$\xspace}

\received{}
\revised{}
\accepted{\today}
\submitjournal{ApJ}

\shorttitle{}
\shortauthors{Moretti et al.}
\graphicspath{{./}{figures/}}

\begin{document}

\title{Observing ram pressure at work in intermediate redshift clusters with MUSE: the case of Abell 2744 and Abell 370}

\correspondingauthor{Alessia Moretti}
\email{alessia.moretti@inaf.it}

\author[0000-0002-1688-482X]{Alessia Moretti}
\affiliation{INAF-Padova Astronomical Observatory, Vicolo dell'Osservatorio 5, I-35122 Padova, Italy}

\author[0000-0002-3585-866X]{Mario Radovich}
\affiliation{INAF-Padova Astronomical Observatory, Vicolo dell'Osservatorio 5, I-35122 Padova, Italy}

\author[0000-0001-8751-8360]{Bianca M. Poggianti}
\affiliation{INAF-Padova Astronomical Observatory, Vicolo dell'Osservatorio 5, I-35122 Padova, Italy}

\author[0000-0003-0980-1499]{Benedetta Vulcani}
\affiliation{INAF-Padova Astronomical Observatory, Vicolo dell'Osservatorio 5, I-35122 Padova, Italy}

\author[0000-0002-7296-9780]{Marco Gullieuszik}
\affiliation{INAF-Padova Astronomical Observatory, Vicolo dell'Osservatorio 5, I-35122 Padova, Italy}

\author[0000-0002-4382-8081]{Ariel Werle}
\affiliation{INAF-Padova Astronomical Observatory, Vicolo dell'Osservatorio 5, I-35122 Padova, Italy}

\author[0000-0002-6179-8007]{Callum Bellhouse}
\affiliation{INAF-Padova Astronomical Observatory, Vicolo dell'Osservatorio 5, I-35122 Padova, Italy}

\author[0000-0002-8372-3428]{Cecilia Bacchini}
\affiliation{INAF-Padova Astronomical Observatory, Vicolo dell'Osservatorio 5, I-35122 Padova, Italy}

\author[0000-0002-7042-1965]{Jacopo Fritz}
\affiliation{Instituto de Radioastronomia y Astrofisica, UNAM, Campus Morelia, AP 3-72, CP 58089, Mexico}

\author[0000-0001-9976-1237]{Genevieve Soucail}
\affiliation{Institut de Recherche en Astrophysique et Planétologie (IRAP), Université de Toulouse, CNRS, UPS, CNES,14 Av. Edouard Belin, 31400 Toulouse, France}

\author[0000-0001-5492-1049]{Johan Richard}
\affiliation{Univ. Lyon, Univ Lyon1, ENS de Lyon, CNRS, Centre de Recherche Astrophysique de Lyon UMR5574, 69230 Saint-Genis-Laval,France}

\author[0000-0001-9575-331X]{Andrea Franchetto}
\affiliation{Dipartimento di Fisica e Astronomia “Galileo Galilei”, Universit\'a di Padova, vicolo dell’Osservatorio 3, IT-35122, Padova, Italy}
\affiliation{INAF-Padova Astronomical Observatory, Vicolo dell'Osservatorio 5, I-35122 Padova, Italy}

\author[0000-0002-8238-9210]{Neven Tomi\v{c}i\'{c}}
\affiliation{INAF-Padova Astronomical Observatory, Vicolo dell'Osservatorio 5, I-35122 Padova, Italy}

\author[0000-0002-0838-6580]{Alessandro Omizzolo}
\affiliation{Vatican Observatory, Vatican City, Vatican State}

\begin{abstract}
Ram pressure stripping has been proven to be effective in shaping galaxy properties in dense environments at low redshift. 
The availability of MUSE observations of a sample of distant (z$\sim 0.3-0.5$) clusters  has allowed to search for galaxies subject to this phenomenon 
at significant lookback times.
In this paper we describe how we discovered and characterized 13 ram-pressure stripped galaxies in the central regions of two intermediate redshift (z$\sim$0.3-0.4) clusters, A2744 and A370, using the MUSE spectrograph. Emission line properties as well as stellar features have been analyzed to infer the presence of this gas--only stripping mechanism, that produces spectacular ionized gas tails (\Ha and even more astonishing [OII](3727,3729) departing from the main galaxy body. The inner regions of these two clusters reveal 
the predominance 
of such galaxies among blue star-forming cluster members, 
suggesting that ram-pressure stripping was even more effective at intermediate redshift than in today's Universe.
Interestingly, the resolved [OII]/\Ha line ratio in the stripped tails is exceptionally high compared to that in the disks of these galaxies, (which is comparable to that in normal low-z galaxies),
suggesting lower gas densities and/or an interaction with the hot surrounding ICM. 

\end{abstract}



\section{Introduction} \label{sec:intro}
Clusters of galaxies are among
the densest regions of the Universe, and are characterized by a peculiar morphological mix of their galaxy population that results in the prevalence of early type galaxies over the late type ones \citep[see][for the nearest clusters]{dressler80,Fasano2015}.
Environmental effects 
are playing a role in shaping galaxy properties, and in particular in quenching the star formation while galaxies approach the cluster center. 
Among the various mechanisms suggested to produce this effect, ram-pressure stripping due to the interaction of the galaxies with the dense and hot Intra Cluster Medium (ICM) permeating the cluster results in the stripping of the neutral gas from the galaxy disks and the subsequent halting of the star formation \citep[][]{GG72}, that leads to the 
morphological transformation of the galaxies.
It has been argued in fact that moving from higher to lower redshift the population of late type galaxies in clusters gets transformed in nowadays S0 galaxies \citep[][]{Dressler+1997,Poggianti+2009}.
Moreover, before being quenched by ram--pressure, galaxies 
may or may not experience an increased episode of star formation \citep[][ and reference therein]{Poggianti+2016, Vulcani+2018,GASPXXIV, Roberts2020RamFormation, Vollmer+2012, Gavazzi+2013,Boselli+2014}.
When their star formation is abruptly stopped, their spectra
show a lack of nebular emission lines and strong Balmer absorption lines \citep[][]{Dressler+Gunn1992,Gavazzi+2010,Gavazzi+2013}. 
Galaxies with such spectra are generally named Post Star Burst galaxies (PSB), irrespective of the fact that they experienced a significant burst or a normal level of star formation before the quenching. While the same type of spectrum can originate from different quenching mechanisms \citep{Pawlik+2019}, ram pressure is by far the most likely responsible in galaxy clusters.

First evidence of ram pressure stripping in action  comes from the hunting of neutral gas stripped tails, that has been possible initially only for the Virgo  and Coma nearby clusters \citep[][]{
Cayatte+1990,
Kenney2004,Chung+2007,Chung2009,Bravo-Alfaro+2002,vanGorkom2004}.
More recently, neutral gas tails have been observed also in slightly more distant (z$\sim 0.05$) ram pressure stripped galaxies within the GASP survey \citep[][]{GASPXVII, GASPXXV, GASPXXVI, Luber+2021}.

Nearby cluster galaxies were also inspected to look for ionized gas tails \citep[see among the others][]{Yagi2010,Fossati2012, Boselli2016} and the pioneristic work by \citealt{Gavazzi2001}.
In fact, the characterization of ram pressure stripped galaxies in low redshift clusters and the consequent physical understanding of the RPS mechanism
have enormously benefited from the advent of optical IFUs, which has allowed the characterization of both the stellar component and of the ionized gas at the same time \citep[][]{Merluzzi2013,Fumagalli2014,Fossati2016}.
Beautiful examples of galaxies caught in the act of losing their gas that shine at \Ha wavelength due to the ongoing star formation are those presented by the GASP survey \citep[e. g. ][]{GASPI, GASPII, GASPIII, GASPIV, GASPV, GASPXIII}, that make use of the MUSE IFU.

At high redshift (z$\sim0.6-1$)
 clusters are in a fast mass-growing phase 
(a 10$^{14}$ \Msun cluster doubles its mass between redshift 0.6 and redshift 0, \citealt{Poggianti+2006})
and there are even fewer relaxed systems.
Therefore, one might expect to find even higher fractions of ram-pressure stripped galaxies as they infall toward the cluster center. However,
there are neither clear predictions, nor observational answers yet, regarding
the preponderance of ram pressure stripping
in distant clusters, also with respect to other mechanisms, nor about the role of ram pressure in the quenching and morphological transformations that are known to have involved a large fraction of today's clusters galaxies since $z\sim 1$ \citep[][]{Pallero+2019,Pintos-Castro+2019}. Open questions include how ram pressure in distant clusters might depend on potentially different ISM and ICM conditions, whether we can still detect evidence for ram pressure stripping at higher redshift, and up to what redshift, and how many of the galaxies in clusters today went through a significant ram pressure stripping phase. 

The ideal redshift range to explore (z$\sim$1) is beyond the reach of today's spectroscopic capabilities, as the spatial resolution becomes too low  to disentangle stellar and gas kinematics. Future instrumentation will help in this respect \citep[][]{mavis}.

From the observational point of view, there are a few detections of ram pressure stripped galaxies in not-local clusters, with
the first studies using a combination of HST and Radio/multiwavelength data \citep{Owen+2006}, together with optical spectroscopic confirmation \citep{Cortese+2007}
 and, more recently, deep HST imaging only \citep[][]{Owers+2012,Rawle+2014,Ebeling+2014,McPartland+2016,Durret+2021}, that has led to the identification of ram pressure stripping candidates characterized by star forming tails emitting in the bluer HST bands.
The HST slitless spectroscopy of the GLASS survey \citep{Treu+2015,Vulcani+2015_glass_V, Vulcani+2016_glass_VII} has allowed the characterization of the \Ha emission in 10 clusters at redshift $\sim 0.3-0.7$: this suggests the possible presence of ram-pressure stripped galaxies correlated with the hot, X-ray emitting Inter Galactic Medium (IGM) of unrelaxed clusters, as also supported by dedicated numerical simulations \citep{Vulcani+2017_glass_VIII}.

The first spectroscopic evidence of ram-pressure events in a z=0.7 cluster, detected using the MUSE IFU spectrograph, is reported in \cite{Boselli+2019} where two long tails of ionized gas departing from two galaxies close to the cluster center have been revealed through their [OII] emission.
These two galaxies show a very large line-of-sight velocity with respect to the host cluster, and still display a tail in the plane of the sky, due to an edge-on stripping.

Similarly, using the KCWI IFU data \citet{Kalita2019} resolved the gaseous and stellar kinematics of a jellyfish galaxy in the intermediate redshift cluster A1758 (z$\sim 0.3$).
A few galaxies in the CIZA J2242.8+5301 (a.k.a. the Sausage) cluster (z $\sim$ 0.2) analyzed by means of GMOS IFU spectroscopy have  been suspected to be ram-pressure stripped galaxies \citep{Stroe+2020}.

For distant clusters, IFUs are not only playing an important role in establishing the real nature of peculiar objects,
but can be a very efficient tool to identify ram pressure stripped galaxies: already at z=0.3, a single MUSE pointing is able to cover the central regions of clusters and study their whole galactic population.
In this paper we will present the  characterization of 13 galaxies with ionized gas tails due to ongoing ram-pressure stripping in two massive intermediate redshift ($\sim 0.3-0.4$) clusters made possible by the MUSE Guaranteed Time Observations (GTO) data \citep[][]{Bacon+2017,Richard+2021}.
This is part of an ongoing, systematic search of tailed ram pressure stripped galaxies in a large sample of intermediate redshift clusters (0.3-0.6). 
Sec. \ref{sec:sample} presents the main characteristics of the two clusters and the criteria that we used to identify candidate ram-pressure stripped galaxies using the MUSE data; Sec. \ref{sec:data_analysis} deals with the description of the procedures that we have used to measure the ionized gas emission lines, as well as the stellar kinematics, and also explains how we defined the galaxy disks. It also uses one of the selected galaxies to illustrate the procedure, presenting the main plots that we produced for all the analyzed galaxies in the sample (shown in the Appendix). Sec. \ref{sec:rps} presents a characterization of the galaxies and their tails. Sec. \ref{sec:o2ha} shows the peculiar line ratios of the [OII] and \Ha lines that we find in these galaxies. We finally draw our conclusions in Sec. \ref{sec:conclusions}.

We will use in the paper a standard cosmology of (H$_0$=70 \Hunit, $\Omega_M$=0.3, $\Omega_\lambda$=0.7).

\section{Galaxy and cluster sample}\label{sec:sample}

Thanks to the MUSE-GTO program and other lensing studies of distant clusters, deep MUSE data are now available for many clusters at intermediate redshift, where MUSE single pointings or mosaics covering the cluster centers have been observed with exposure times that go from $\sim2$ to $\sim$15 hours (effective).
The clusters covered by MUSE observations are extracted from the MAssive Clusters Survey \citep[MACS,][]{MACS}, Frontier Fields \citep[FFs,][]{FF}, Grism Lens-Amplified Survey from Space \citep[GLASS,][]{GLASS} and Cluster Lensing And Supernova survey with Hubble \citep[CLASH,][]{CLASH} programmes. 
The 5$\sigma$ emission line detection limit for a point like source that has been achieved by these observations is in the range between $(0.77–1.5)\times 10^{-18}$ erg s$^{-1}$ cm$^{-2}$ at 7000 \AA.
The data analysis of the entire sample of clusters is described in \cite{Richard+2021}, where also the full redshift catalogs are given.

Among the surveyed fields we selected 12 clusters, in a redshift range between $\sim 0.3$ and $\sim 0.5$.
We focus this first paper of the series on the detection of peculiar galaxies in the two clusters A2744 and A370, the first two clusters that have been observed within the MUSE GTO program \citep{Mahler+2018,Lagattuta+2019}. The analysis of the ram-pressure stripped galaxies in the remaining sample of 10 clusters will be presented in future papers.
The radius of the region covered by MUSE observations corresponds to $\sim 250-330$ kpc, depending on the cluster's redshift, roughly matching the inner $\sim 0.1-0.15 R_{200}$.

Both clusters are massive systems, and contain a large number of  spectroscopically confirmed members within the MUSE observed FOV (153 and 227 for A2744 and A370, respectively, if selecting those with a high confidence on the redshift determination from \citealt{Richard+2021}).
A2744 (also known as the Pandora cluster) has a redshift of 0.306 \citep{Owers2011} and a velocity dispersion of  $\sim 1800$ \kms \citep{Boschin2006}, but shows a double peak behaviour in the redshift distribution, hinting toward a complex system. Its mass within 1.3 Mpc is $\sim 2 \times 10^{15}$ \Msun \citep[][]{Jauzac+2016}.

A370 is a 
comparably massive cluster ($8 \times 10^{14}$ \Msun within 500 kpc from \citealt{Lagattuta+2017}), with a velocity dispersion of $\sim 1300$ \kms and a $R_{200}=2.57$ Mpc \citep{Lah2009}. Its redshift is 0.373, i.e. it is more distant than A2744.
A370 is known to possess at least two substructures, each of them characterized by a velocity dispersion of $\sim 850$ \kms \citep{Kneib+1993}, and a recent analysis \citep{Richard+2010} confirms  that this is probably due to a recent merging event.
The presence of a diffuse faint radio halo in the southern part also points towards this conclusion \citep{Xie+2020}.
The analysis of the HI emitting galaxies in A370 by \citet{Lah2009} shows, indeed, an asymmetric redshift distribution with a small tail at lower redshift.

Table \ref{tab:clusters} gives the main cluster features, together with the scale that we will use in what follows.

\begin{table}[t]
    \centering
    \caption{Cluster properties: A370 data taken from \citealt{Lah2009} except the total mass from \citealt{Lagattuta+2017} and the X--ray luminosity from \citealt{Morandi+2007}, A2744 coordinates and redshift from \citealt{Owers2011}, A2744 velocity dispersion and r$_{\rm 200}$ from \citealt{Boschin2006} and total mass from \citealt{Jauzac+2016}.}
    \small
    \begin{tabular}{|l|l|l|}
    \hline
    \hline
    & A2744 & A370\\
    \hline
RA$_{\rm bcg}$   & 00:14:20.738 & 02:39:52.90 \\
DEC$_{\rm bcg}$  & -30:23:59.90 & -01:34:37.5 \\
z$_{\rm centre}$ & 0.306 & 0.373 \\
$\sigma_v$ (\kms)  & 1767 & 1263 \\
R$_{\rm 200}$ (Mpc)& 2.00 & 2.57 \\
Mass (M$_\odot$) & $2.3 \times 10^{15}$ (R$<$1.3 Mpc) &  $8 \times 10^{14}$ (R$<$500 kpc) \\
L$_X$ (erg $s^{-1}$) & 3.1 $\times 10^{45}$ & 1.1 $\times 10^{45}$ \\
scale (kpc/\arcsec) & 4.19 & 4.73\\
 \hline
    \end{tabular}
    \label{tab:clusters}
\end{table}

Our aim is to identify, using MUSE datacubes, two types of galaxies: star-forming emission-line galaxies with clear signatures of ram pressure stripping (hereafter RPS) and passive galaxies that were recently quenched which can be post-starburst or post-starforming (hereafter PSB, for simplicity). According to this operative definition there is no overlap between the two samples, as only RPS galaxies do show more or less evident signatures of stripped gas tails.

\begin{figure*}
    \centering
\includegraphics[width=0.45\textwidth]{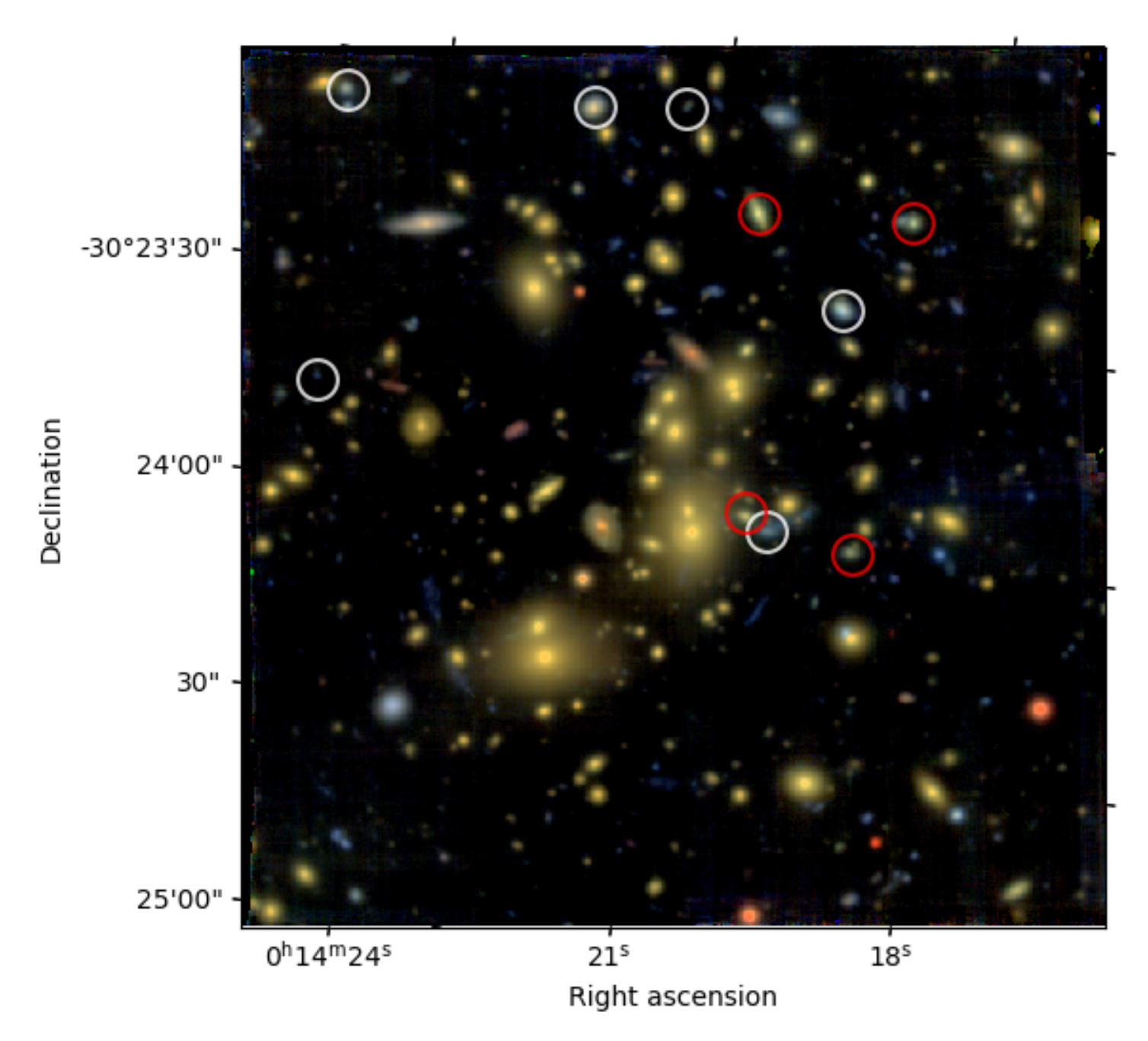}
\includegraphics[width=0.45\textwidth]{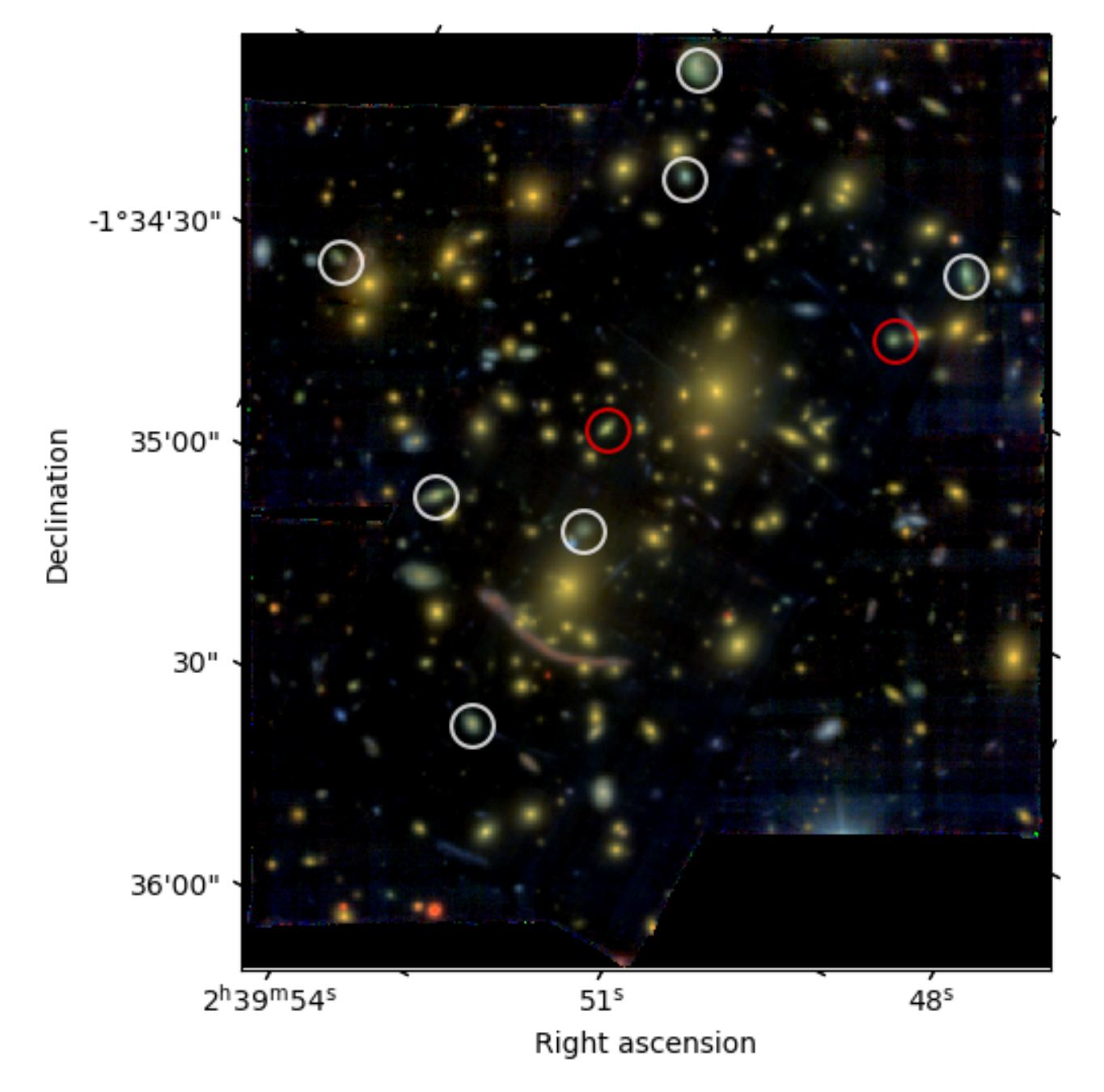}
    \caption{gri images of the central $\sim 2 \arcmin \times 2 \arcmin$  of A2744 (left) and A370 (right) clusters from the MUSE datacubes. White/red circles show the positions of our RPS/PSB galaxies.}
    \label{fig:RGB_clusters}
\end{figure*}
With this aim, three of us (BP, BV and MG) inspected the MUSE datacubes (here shown in Fig.\ref{fig:RGB_clusters}) and the HST RGB images (F435W+F606W+F814W). RPS galaxies were identified searching for extraplanar, unilateral tails/debris with emission lines in the MUSE datacubes and/or unilateral tails/debris from HST images that were confirmed to belong to the galaxy from the MUSE redshifts. PSB galaxies were chosen to be generally lacking emission but to display evident strong Balmer lines in absorption from the MUSE data.
All galaxies in the MUSE+HST data were searched in particular for these signatures.
Probable mergers and weak RPS cases have also been tagged but are not included in this paper. 

Our final sample comprises 10 RPS/PSB galaxies in A2744 and 9 RPS/PSB galaxies in A370 (one of which possibly belonging to a secondary substructure). 
We recall here that MUSE observations cover only the innermost regions of clusters (out to $\sim 0.15 R_{200}$), and more RPS galaxies could be present further out.
We focus in this paper on the spectroscopic characterization of the RPS galaxies only, while we leave the detailed analysis of the PSB galaxies to a separate study (Werle+, in prep.).

\begin{figure*}
    \centering
\includegraphics[width=0.9\textwidth]{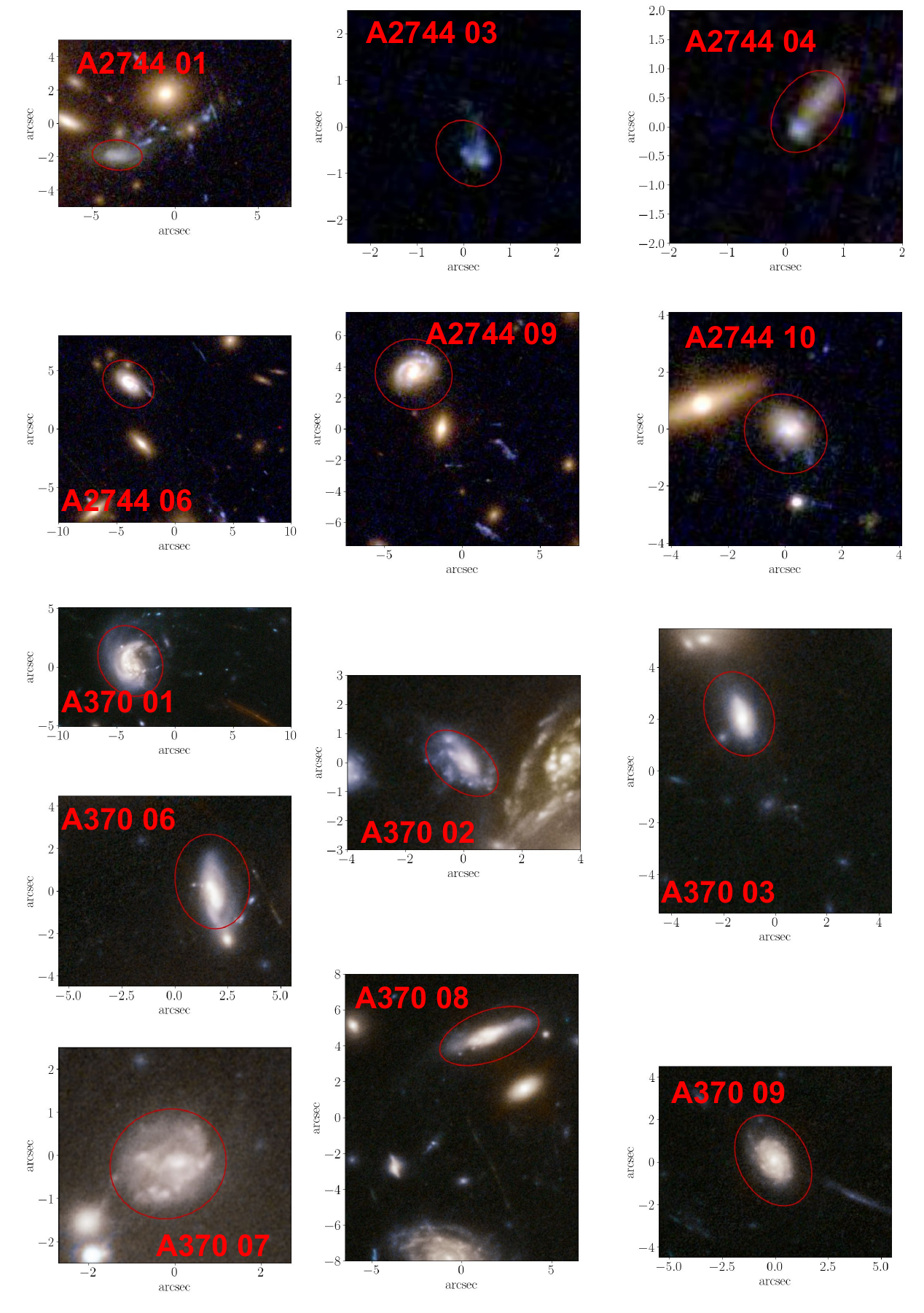}
    \caption{RGB images of the sample RPS galaxies in A2744 and A370 obtained from HST data (F435W, F606W, F814W). The red ellipses show the extent of the stellar disk (see Sec.\ref{sec:data_analysis}).}
    \label{fig:RGB}
\end{figure*}

Fig. \ref{fig:RGB} shows the RGB images of the RPS galaxies that we have selected in the two clusters, constructed 
using the HST filters.

Overplotted on the RGB images we show the galaxy disks as traced by their stellar emission (see Sec. \ref{sec:data_analysis}).
While in some cases the HST images are able to highlight the blue clumps related to the star formation in the tails (for example in A2744 01, A2744 03, A2744 09 and A370 03),  in other cases the visual inspection of the beautiful HST images does not always reveal the disturbed nature of the galaxies, therefore only the MUSE line emission analysis can clearly characterize them as RPS (or PSB).

In fact, among the selected galaxies only A2744 01 has already been identified as jellyfish galaxy by \citealt{Owers+2012}, where it was named ``central'', but was lacking a spectroscopic confirmation.
The other galaxies were missed by previous studies analyzing the cluster galaxies population by means of both HST imaging \citep{Rawle+2014} and HST slitless spectroscopy \citep{Vulcani+2016_glass_VII}. In particular the galaxies A2744 09, A2744 06 and A370 06 were classified as mergers, while A370 03 and A370 08 were classified as undisturbed galaxies using the GLASS spectra.

\section{Data analysis}\label{sec:data_analysis}
From the MUSE datacube, we extracted a cubelet containing each selected galaxy over an area encompassing all the interesting features that we detected by visually inspecting the cube.
We then corrected the cube for Galactic extinction using the values estimated at the galaxy position from \citet{mw_ext} and the extinction law by \cite{CCM1989}, and we smoothed it with a 5x5 pixel boxcar kernel, corresponding to the typical PSF size of $\sim$1\arcsec.
The spectral resolution of the observations varies from R$\sim$2000 to R$\sim$4000, with a spectral range between 4750 and 9350 \AA. The corresponding velocity resolution is $75-150$ \kms sampled with $\sim 55$ \kms pixels.
On the spatially smoothed cube we then performed a first run of emission lines spectral fitting using a purposely-devised software described below, and a determination of the galaxy stellar kinematics using the pPXF software \citep{CE2004} binning the data in Voronoi bins with S/N=5. With the redshifts derived from emission and absorption lines we then used the SINOPSIS code \citep{GASPIII} to extract the best fitting mix of Single Stellar Populations (SSP) that best reproduces the stellar features and produced a second version of the MUSE cube containing only the gas emission lines.
SSPs used by the SINOPSIS code are given by S. Charlot \& G. Bruzual (in prep.) and adopt a \citet{Chabrier2003} IMF with stellar masses between 0.1 and 100 \Msun and metallicities between 0.004 and 0.04. 
On the emission only cubes we measured again the emission lines fluxes and finally corrected them for internal extinction using the observed value of the Balmer decrement and the \citet{CCM1989} extinction law.

In order to measure the emission lines, we developed  our own software ({\it HIGHELF}, Radovich et al., in prep.), that is able to fit the entire set of emission lines included in the observed redshift range.
{\it HIGHELF}  is based on the LMFIT Python library (https://lmfit-py.readthedocs.io/), and is designed to fit a user-defined set of emission lines with either one or two Gaussian components. It starts by computing the zero-moment of the cube, i.e. the integral of the flux within the entire observed wavelength range, and the noise given by its variance. This constitutes the initial Signal--to--Noise Ratio (SNR) of each spaxel.
Then, for each spaxel an initial estimate  of the galaxy redshift is refined by an iterative procedure where different subsets of lines are fitted simultaneously, and the optimal result is found. Finally, each emission line is fitted independently 
with the Gaussian mean value (velocity) and standard deviation (velocity dispersion)
as free parameters.
The set of emission lines that are included in the observational setup at the clusters redshift is given in Tab.\ref{tab:linelist}.

\begin{table}[t]
    \centering
    \caption{Emission lines fitted with \it{HIGHELF}.}
    \begin{tabular}{|l|l|}
    \hline
    \hline
    Line & $\lambda [\mathrm{\AA}]$ (air)\\
    \hline
    $[$OII$]$ & 3726.03\\
    $[$OII$]$ & 3728.81\\    
    H$_\delta$ & 4101.74\\
    $[$HeI$]$ & 4026.19 \\
    $[$HeI$]$ & 4143.76\\
    H$_\gamma$ & 4340.47\\
    $[$OIIIa$]$ & 4363.21\\
    $[$HeII$]$ & 4685.71 \\
    H$_\beta$ & 4861.33\\
    $[$OIII$]$ & 4958.91 \\
    $[$OIII$]$ & 5006.84 \\
    $[$OI$]$   & 6300.30 \\
    $[$OI$]$   & 6368.77 \\
    $[$NII$]$  & 6548.05 \\
    H$_\alpha$ & 6562.82\\
    $[$NII$]$   & 6583.46 \\
    $[$SII$]$   & 6716.44 \\
    $[$SII$]$   & 6730.81 \\
 \hline
    \end{tabular}
    \label{tab:linelist}
\end{table}

For the subsequent analysis, we selected spaxels with a SNR (as derived by {\it HIGHELF})$>$2. 
Each measurement was then flagged according to the S/N in the given line, the error in the velocity and in the velocity dispersion, the difference between the velocity/velocity dispersion values with respect to their mean value
(values larger than 3$\sigma$ are flagged).

We also compared the {\it HIGHELF} measurements with those obtained with the KUBEVIZ \citep{Fossati2016} software that is being commonly used for low redshift IFU analysis, and found that the two are in agreement within the errors.
For the lines that were outside the observed range at low redshift we visually checked that the automatic {\it HIGHELF} measurements are correct: in the case of the [OII] (3727,3729) doublet, we also verified that the software was able to deblend it.
To derive the total [OII] flux we fitted both lines in the doublet and then summed together the two contributions.

We corrected all the spaxels according to their Balmer decrement, irrespective of their classification in the BPT diagram \citep[][]{BPT} involving the [NII]$\lambda$6583 line, assuming a dereddened Balmer decrement of \Ha/\Hb=2.86. This value is in principle only appropriate in case B recombination for a T=10000K medium \citep[][]{Osterbrock1989}, and therefore the estimate of A$_V$ may not be correct if other ionization mechanisms (e.g. AGN, shocks) are dominating, instead of photoionization in HII regions.

To define the galaxy boundary and be able to state what is within the galaxy disk and what is the stripped material we developed a procedure based on the MUSE g-band reconstructed image. As a first step, we defined the centre of the galaxy of interest as the centroid of the brightest central region in the map. 
We then identified the surface brightness of the sky background, by masking the galaxy itself and the neighbours, if any. Next, we identified the stellar isophote corresponding to a surface brightness 3$\sigma$ above the measured sky background level. In the case of neighbours, we adjusted the isophotes to remove their contributions. As the resulting isophote is quite jagged for some galaxies, we finally fit an ellipse to the isophote. The resulting contour defines a mask that we use to discriminate the galaxy main body and the ram-pressure stripped tail. In the following, we will refer to the regions within this mask as the galaxy main body, and to the regions outside the mask as tails. 

As ram pressure stripping is able to remove the gas from the galaxy disks, without affecting the stellar component, we used the comparison between the stellar and the gaseous extension to assess whether the observed gaseous tails are due to this physical process.
The detailed maps, as well as a discussion for each single galaxy will be given in the  Appendix for all RPS galaxies.
We discuss in the next section the maps for the A2744 06 galaxy, used as a showcase.

\subsection{A2744 06 as a showcase}
We use as an example here the galaxy A2744 06, that is a clear ram-pressure stripped galaxy.
In particular, 
we would define it a jellyfish galaxy, as it has a long ionized gas tail.
\begin{figure*}
    \centering
   \includegraphics[width=0.85\textwidth]{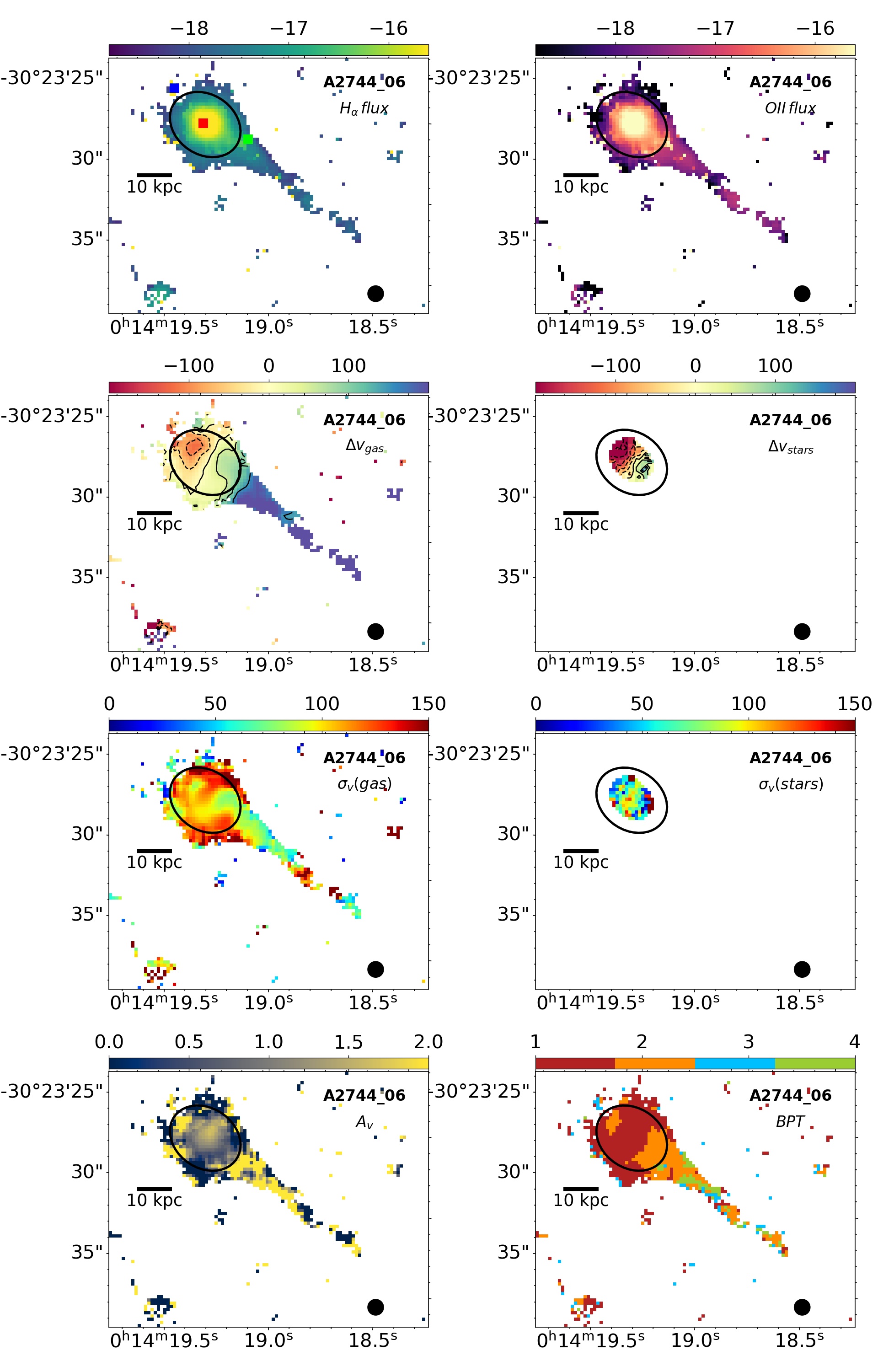}
    \caption{\Ha and [OII] fluxes in erg/cm$^2$/s/arcsec in logarithmic units, \Ha and stellar velocities in \kms, \Ha and stellar velocity dispersions in \kms, A$_V$ and BPT map for A2744 06 (1=Star Formation, 2=Composite, 3=Liner/shocks, 4=AGN). The three colored squares in the top left panel show the position of the regions from which we extracted the spectra shown in Fig. \ref{fig:spectra}.}
    \label{fig:a2744_06}
\end{figure*}
From our analysis of the emission lines in the MUSE emission only (i. e. corrected for the underlying stellar absorption) spectra we estimated the \Ha extinction corrected flux that is shown in the upper left panel of Fig.\ref{fig:a2744_06}, as well as the [OII] flux shown in the right panel. 
\footnote{The emission in the south east part of the map belongs to a galaxy at redshift z=0.2983 (i.e. at a projected velocity difference of $\sim 700$ \kms) at $\sim 60$ kpc from A2744 06, which is highly unlikely to be responsible for the gas tail.}
From the fit of the \Ha emission line we also derived the spatially resolved ionized gas kinematics, while the stellar one has been measured on Voronoi binned regions with S/N$>5$.
The comparison of the two (shown in the left and right panels of the second row in Fig.\ref{fig:a2744_06}) allowed us to infer the signature of the ram pressure, given that the stellar kinematics is not disturbed, as expected if the interaction of the galaxy with the ICM only affects the gaseous component of the disk.
A clear velocity gradient can be seen along the tail that reaches
$\sim$ 300 \kms with respect to the galaxy center: we interpret this as ionized gas left behind along a tail from the galaxy moving not only in the plane of the sky, but also with a non negligible component of the motion along the line of sight.
Indeed, the maximum projected line-of-sight velocity of the gas in the galactic disk is 150-200 km/s (for the receding side, or -150-200 km/s for the approaching side), which is smaller than the line-of-sight velocity of the gas in the tail.

The stellar velocity dispersion is $\sim$ 120 \kms at the galaxy center (not corrected for the beam smearing effect), and the gaseous one is slightly higher. It is instead generally much lower in the tail.

By looking at the two bottom panels  of Fig. \ref{fig:a2744_06} it can be seen that the average A$_V$ within the stellar star forming disk is $\sim 1$ mag, while it seems to increase along the tail. 
Spaxels in the disk are mainly powered by star formation, with the emerging of composite regions toward the tail.

\begin{figure}
    \centering
   \includegraphics[width=0.45\textwidth]{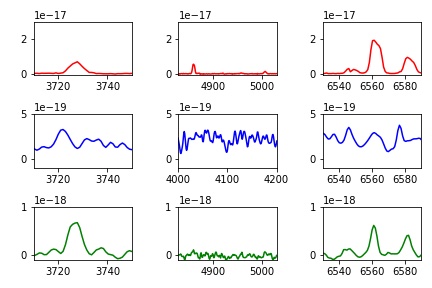}
    \caption{Zoom spectra (in erg/cm$^2$/s/\AA) at restframe wavelength of 3$\times$3 pixels regions extracted from different regions of the A2744 06 galaxy: top row refers to the central spaxels (in red), middle row to the north-east region just outside the stellar disk (in blue), lower panels to a region in the tail (in green). The spectral windows cover the [OII] lines (left), the \Hb and [OIII] lines (middle) and the \Ha and [NII] lines (right) for the central and tail regions, while the wavelength range around \Hd is shown for the PSB region.}
    \label{fig:spectra}
\end{figure}

Fig.\ref{fig:spectra} shows typical spectra of different regions of  A2744 06.
The top row shows three zoomed spectral regions of the emission extracted within an aperture of 3 pixels around the center of the galaxy (also shown as red box in the top left panel of Fig. \ref{fig:a2744_06}): the first panel shows the [OII] lines (here blended), the second panel the \Hb and [OIII] lines and the right panel the region covering the \Ha and the [NII] lines.
In the second row we show the same plots for a region located just outside the stellar disk at the opposite side of the tail (blue box in the top left panel of Fig. \ref{fig:a2744_06}), i.e. where the gas should not be present anymore, if the stripping direction is traced by the tail direction. In fact, all the spectra show the absence (or a very marginal 1$\sigma$ detection) of emission lines. In the middle panel we show in this case the wavelength range around the \Hd absorption, typical signature of a PSB emission.
The last row shows, instead,  the zoomed spectra extracted from a region in the tail (green box in the top left panel of Fig. \ref{fig:a2744_06}), where ionized gas is present and detectable in both \Ha and [OII] lines, albeit with different proportions with respect to the central region, as will be discussed in Sec.\ref{sec:o2ha}.

\section{Ram pressure stripped galaxies in A2744 and A370}\label{sec:rps}

In this section we describe the main characteristics of the galaxies subject to ram-pressure stripping in A2744 and A370. Their position within the cluster is shown in Fig.\ref{fig:RGB_clusters} where RPS/PSB galaxies are the white/red circles, while the detailed analysis of the cluster's dynamical stage and orbital histories will be discussed in a forthcoming paper (Bellhouse+, in prep.).

Table \ref{tab:sample} contains the galaxy main characteristics (MUSE ID, sky coordinates, redshift and HST magnitudes in different filters (F435W, F606W, F814W), as well as our classification as RPS or PSB galaxy), the stellar mass and the approximate projected tail length.

\begin{table*}[t]
    \centering
    \small
    \begin{tabular}{|l|lllcccccrr|}
    \hline
    ID & ID MUSE    & RA    &   DEC & z & F435W & F606W & F814W & Class & M$_\star$ [log]& tail length \\
    \hline
 A2744 01 &	6982	&   00:14:19.920  &	-30:23:58.93  &	0.292	&	23.58 &	22.47 &	22.26  &	RPS &  8.5$\pm$0.3 & 30\\
 A2744 03 &	8164	&	00:14:24.941  &	-30:23:46.45  &	0.303 	&	25.30 &	25.18 &	25.27 &	RPS &  7.9$\pm$0.2 & 4\\
 A2744 04 &	11203	&	00:14:21.400  &	-30:23:02.06  &	0.294 	&	25.98 &	25.47 &	25.27 &	RPS &  8.0$\pm$0.2 & 5\\
 A2744 06 &	10243	&	00:14:19.440  &	-30:23:26.96  &	0.293 	&	21.29 &	20.39 &	19.91 &	RPS & 10.0$\pm$0.1 & 42\\
 A2744 09 &	11259	&	00:14:22.393  &	-30:23:03.65  &	0.296 	&	21.57 &	20.28 &	19.55 &	RPS & 10.6$\pm$0.1 & 47\\
 A2744 10 & 11908	&	00:14:25.056  &	-30:23:05.86  &	0.296 	&	23.03 &	21.97 &	21.45 &	RPS & 9.3$\pm$0.1 & 13\\
 A370 01  &	15715	&	02:39:54.452  &	-01:33:36.41  &	0.374 	&	21.09 &	20.15 &	19.61 &	RPS & 10.6$\pm$0.2 & 35\\
 A370 02  &	9534	&	02:39:56.760  &	-01:34:28.28  &	0.424 	&	23.11 &	22.20 &	21.65 &	RPS &  9.6$\pm$0.2 & 5\\
 A370 03  &	12808	&	02:39:54.044  &	-01:33:52.09  &	0.358	&	22.29 &	21.25 &	20.79 &	RPS &  9.7$\pm$0.2 & 19\\
 A370 06  &	16798	&	02:39:51.031  &	-01:33:45.05  &	0.359	&	21.95 &	20.69 &	20.18 &	RPS & 10.0$\pm$0.2 & 19\\
 A370 07  &	8007	&	02:39:53.266  &	-01:34:47.27  &	0.380	&	22.83 &	21.62 &	20.90 &	RPS &  9.8$\pm$0.2 & 7\\
 A370 08  &	8006	&	02:39:54.759  & -01:34:53.29  &	0.388	&	22.27 &	20.93 &	20.21 &	RPS & 10.1$\pm$0.2 & 52\\
 A370 09  &	5109	&	02:39:53.334  &	-01:35:21.56  &	0.347	&	22.37 &	21.09 &	20.36 &	RPS & 10.0$\pm$0.1 & 9\\
\hline
 A2744 02 &	7230	&	00:14:20.175  &	-30:23:56.77  &	0.320 	&	23.22 &	21.63 &	20.95 &	PSB &  9.8$\pm$0.1 &--\\
 A2744 05 &	12195	&	00:14:18.806  &	-30:23:13.48  &	0.308   &	23.20 &	21.73 &	21.22 &	PSB &  9.4$\pm$0.1 &--\\
 A2744 07 &	11856	&	00:14:20.474  &	-30:23:15.10  &	0.300 	&	21.92 &	20.33 &	19.66 &	PSB & 10.2$\pm$0.1 &--\\
 A2744 08 &	6843	&	00:14:18.976  &	-30:24:00.29  &	0.306	&	24.16 &	22.63 &	22.05 &	PSB &  8.9$\pm$0.1 &--\\ 
 A370 04  &	12305	&	02:39:51.365  & -01:33:59.05  &	0.361	&	23.07 &	21.61 &	21.06 &	PSB &  9.4$\pm$0.2 &--\\
 A370 05  &	9825	&	02:39:53.542  &	-01:34:31.81  &	0.390	&	22.92 &	21.28 &	20.55 &	PSB &  9.9$\pm$0.1 &--\\
 \hline
    \end{tabular}
    \caption{Ram Pressure Stripped and Post-Starburst galaxies in the central regions of A2744 and A370. For each galaxy we give the MUSE ID from \citealt{Richard+2021}, the galaxy coordinates, its redshift and the HST magnitudes from \citealt{Richard+2021}. We then list the classification given in this work, the disk stellar mass in $\log(M/M_{\odot})$ derived from the spectral fitting using the \citealt{Chabrier2003} IMF and the approximate tail length in kpc.  
    }
    \label{tab:sample}
\end{table*}

\begin{figure}
    \centering
    \includegraphics[width=0.45\textwidth]{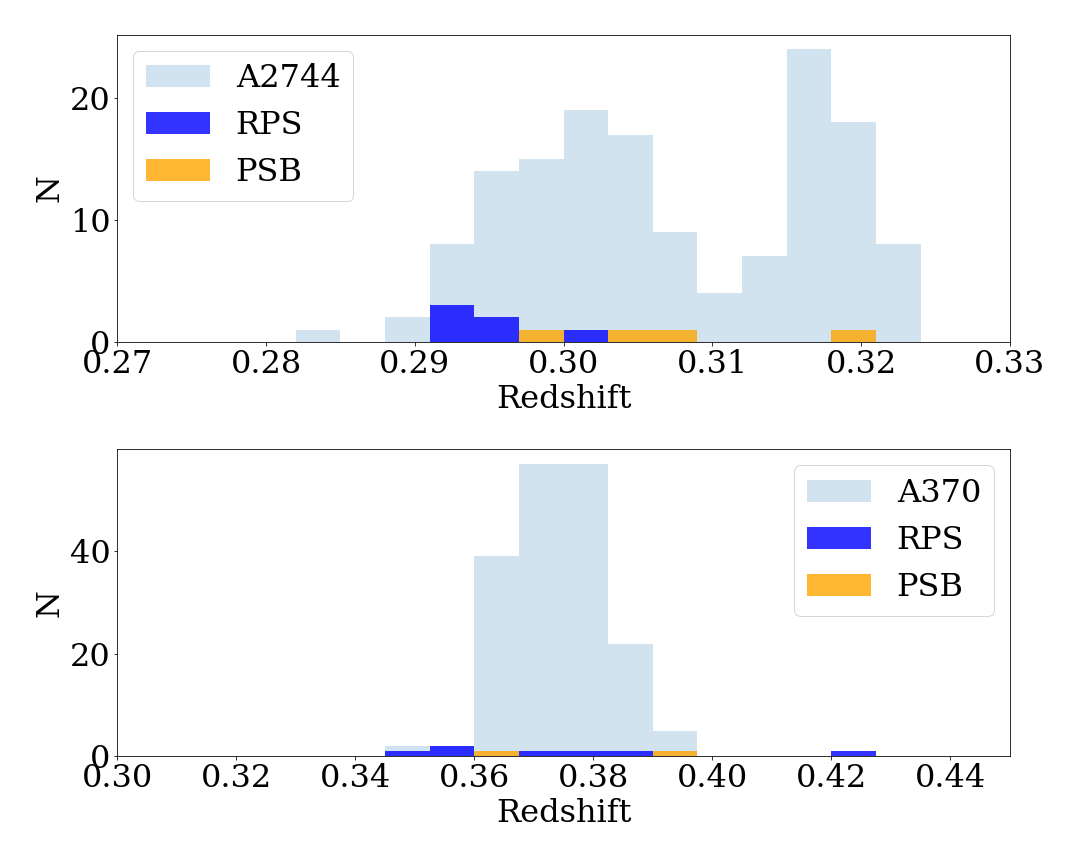}
    \caption{Redshift distribution of the A2744 (upper) and A370 (lower) cluster members with superimposed the distribution of those classified as RPS (blue) or PSB (orange).}
    \label{fig:histoZ}
\end{figure}

The redshift distribution  of the target cluster galaxies is shown in Fig.\ref{fig:histoZ} as light blue histograms.
Two outliers are clearly visible, A2744 02 (classified as PSB) and A370 02 (RPS) that seems to belong to a possible sub-structure. 
It is therefore possible that gas stripping events alter galaxy gas also in infalling substructures.
In fact, an increased fraction of such disturbed galaxies has been proposed also by \citealt{Owers+2012} and  \citealt{EbelingKalita2019} in merging clusters.

\begin{figure}
    \centering
    \includegraphics[width=0.45\textwidth]{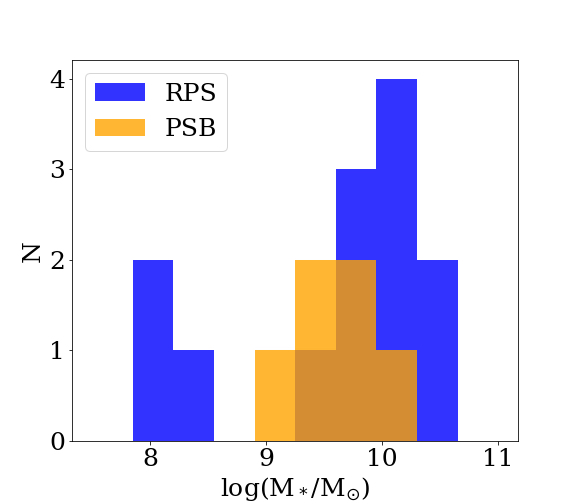}
    \caption{Stellar mass distribution of the galaxies classified as RPS (blue) or PSB (orange).}
    \label{fig:histomass}
\end{figure}

Using the SINOPSIS spectral fitting of the MUSE data within the regions that define galaxy disks, we derived the galaxy stellar masses that are shown in Fig. \ref{fig:histomass} by summing the mass contribution from all the spaxels.
The range of galaxy's masses  where we can infer the presence of the ram--pressure stripping goes from $\sim 8$ to $10.6$ in $\log(M/M_{\odot})$, while PSB galaxies cover a smaller range of stellar masses.
We note, though, that the two galaxies with the lowest stellar masses (A2744 02 and A2744 03) have a low S/N, thus hampering a clear measurement of the stellar kinematics and their classification as stripping galaxies is solely based on the visual inspection of the emission line morphologies.
The third galaxy having a low mass is A2744 01, with $\log(M_{\star})=8.5$, which is a clear example of a galaxy without any ionized gas left in the disk but with a clear tail of gas emission.

The range of SFR within the disks of our sample, as derived from the dust-corrected, absorption-corrected \Ha emission goes from 0 (for PSB galaxies) to 73 \Msun/yr (A370 01) (Vulcani et al., in prep). 

As shown in the spatially resolved maps (in the Appendix) illustrating the resolved properties of the RPS galaxies, all of them show clear emission in the tails, that is visible both from the \Ha emission, and even more clearly using the [OII] line as tracer.

As in several low-z ram-pressure stripped galaxies \citep[][]{Fumagalli2014,GASPI,GASPIV,GASPX,GASPXXII}
the stripped gas maintains the rotation it had in the disk in some cases (A2744 09, A370 01, A370 09), while in other cases the tail velocity reaches very high values, suggesting that galaxies are also moving fast along the line of sight (A2744 06, A370 03).
When long tails (up to $\sim 50$ kpc) are visible, they tend to shrink in size moving away from the disk.
The approximate projected tail length goes from 9 to $\sim 50$ kpc, and is also given in Tab.\ref{tab:sample}.
Generally speaking, galaxy disks are dominated by SF spaxels, according to their positioning in the BPT diagram involving the [NII] line and the demarcation line by \citet{Kauffmann2003}. Composite regions (using the \citealt{Kewley2001} line) are also visible in some galaxies, close to the developed tail. Tails are generally composite regions, with the noticeable exception of A370 06 that seems dominated by AGN-like emission suggesting the presence of an ionization cone (see Fig.\ref{fig:a370_03_06} in the Appendix). 

Similar line ratios 
have been found in low redshift cluster galaxies and often interpreted as due to shocks \citep[][]{Fossati2016,Consolandi+2017}. However, a different origin involving the ionization from a warm plasma surrounding the tails has been suggested recently \citep[][]{GASPXXIII,GASPXXXIV,Ge+2021}.

For what concerns the presence of AGNs, our analysis of the line ratios only reveals clearly two 
AGN candidates (see Appendix): A370 02, with signs of an ionization cone, and A370 06, with a more extended central region classified as powered by an AGN according to the BPT diagram. A2744 09 also shows signs of a possible ionization cone, and is among the most massive RPS in our sample. Overall 2(3) over the 13 RPS galaxies appear to possess an AGN, representing a fraction of 15\%(23\%). Two AGNs are also found in our PSB sample, as discussed in Werle et al. (in prep). The fraction of AGN among low redshift GASP jellyfishes amounts to 24\% when considering the entire mass range, but we can not simply compare the two fractions, as the RPS galaxies mass distribution of the two samples are quite different, and the galaxy mass plays an important role (Peluso et al., submitted). When larger samples of RPS galaxies in distant clusters will become available it will be possible to investigate the incidence of AGNs as a function of galaxy mass and the evolution of the AGN fraction among jellyfish galaxies.

\subsection{Overall incidence of RPS/PSB galaxies}
\begin{figure}
    \centering
    \includegraphics[width=0.45\textwidth]{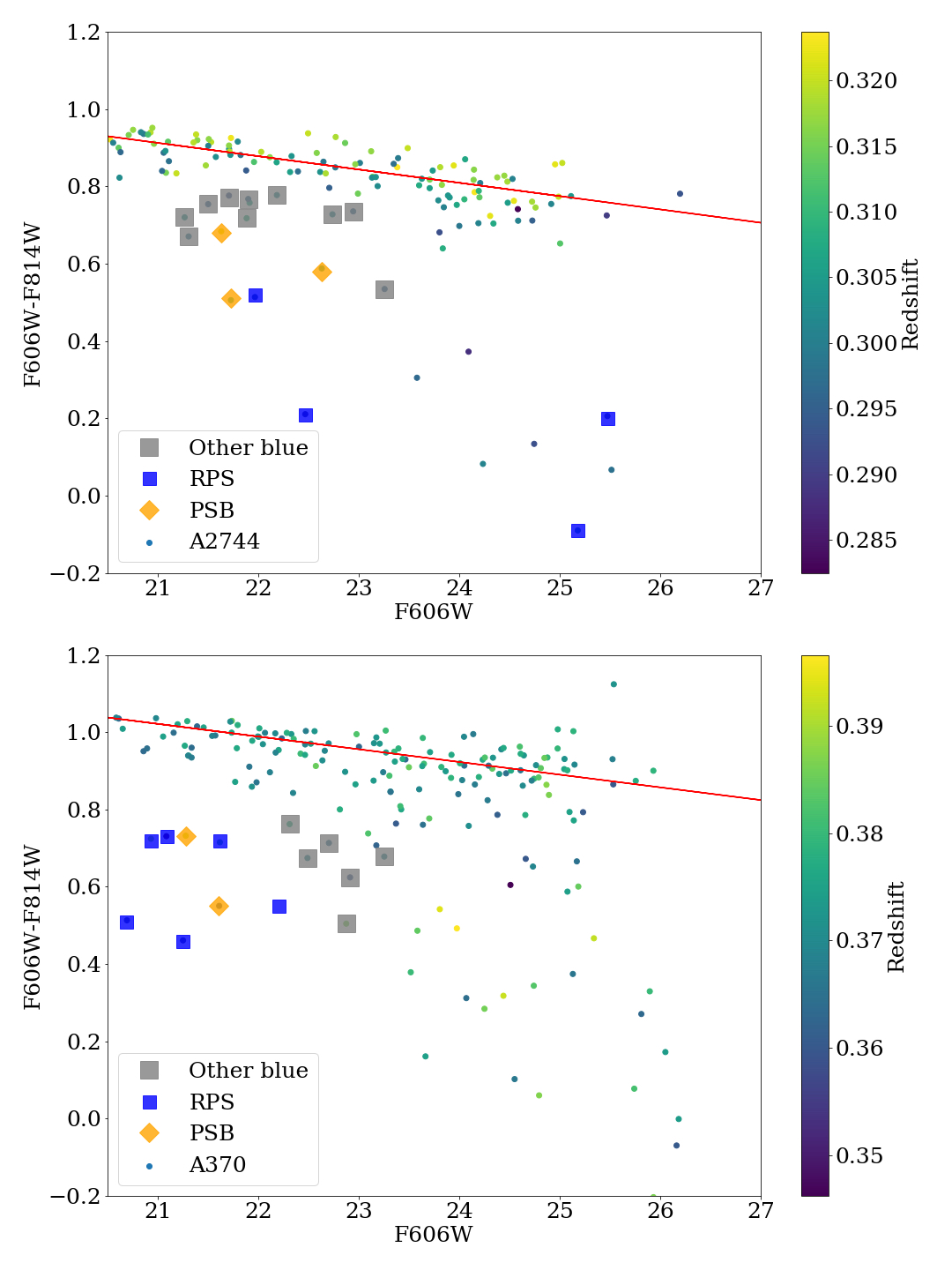}
    \caption{Color-magnitude diagrams of the cluster galaxies 
    in the two clusters: A2744 (top panel) and A370 (bottom panel), color coded according to the redshift. The continuous lines are the 2$\sigma$-clipped fit to the data up to F606W=26, and highlight the clusters red sequence. Ram-pressure stripped galaxies are shown as blue squares, post-starburst as orange diamonds.}
    \label{fig:cm}
\end{figure}

In order to evaluate the importance of ram pressure stripping for the general population of star forming objects in the cluster central regions, we 
analyzed in detail the other blue galaxies within the MUSE fields.
Fig. \ref{fig:cm} 
shows the HST color magnitude diagram 
(F606W, F814W) of the galaxies in each cluster with a spectroscopic redshift within the cluster redshift range given in \citet{Richard+2021}, with superimposed the red sequence derived by performing a 2$\sigma$ clipping fit to galaxies brighter than F606W=26 mag and weighting the magnitudes with the corresponding photometric error. 
We also highlight in the Figure our RPS and PSB galaxies with different symbols: blue squares represent 
RPSs, while yellow diamonds are PSBs.
At magnitudes brighter than F606W=23.5 and colors bluer than (F606W-F814W)$<$0.8, where the great majority of RPSs and PSBs lie (16 of them), there are other 17 galaxies which are neither RPSs nor PSBs shown as grey squares in the Figure. We inspected their MUSE datacubes to study their characteristics that are also summarized in Tab. \ref{tab:control}.
\begin{table*}[t]
    \centering
    \small
    \begin{tabular}{|l|llcccr|}
    \hline
    ID MUSE    & RA    &   DEC & z &  F606W & F814W & type\\
    \hline
A2744 10239	& 00:14:22.359	& -30:23:25.254	& 0.301	& 22.74	& 22.01 & 	faint Ha\\
A2744 10703	& 00:14:23.102	& -30:23:18.65	& 0.297	& 22.94	& 22.21 & 	k\\
A2744 11303	& 00:14:20.092	& -30:23:04.828	& 0.303	& 21.27	& 20.55 & 	k\\
A2744 11655	& 00:14:17.696	& -30:23:09.527	& 0.297	& 21.90	& 21.13 & 	k\\
A2744 12079	& 00:14:22.434	& -30:23:15.345	& 0.296	& 23.25	& 22.72 & 	k\\
A2744 2768	& 00:14:20.570	& -30:24:50.308	& 0.303	& 22.19	& 21.41 & 	k\\
A2744 3328	& 00:14:16.702	& -30:24:43.780	& 0.299	& 21.31	& 20.64 & 	k\\
A2744 9710	& 00:14:20.396	& -30:23:34.346	& 0.296	& 21.50	& 20.75 & 	k\\
A2744 9876	& 00:14:19.289	& -30:23:31.933	& 0.294	& 21.71	& 20.94 & 	k\\
A370 4453	& 02:39:54.591	& -01:35:24.933	& 0.364	& 22.70	& 21.91 & 	k\\
A370 5339	& 02:39:54.047	& -01:35:16.526	& 0.378	& 22.31	& 21.55 & 	k/k+a\\
A370 7653	& 02:39:56.466	& -01:34:49.71	& 0.386	& 22.88	& 22.37 & 	faint k+a\\
\hline
A2744 11651	& 00:14:17.607	& -30:23:10.771	& 0.308	& 21.88 & 21.17	& merger	\\
A370 13663	& 02:39:54.881 	& -01:33:42.781 & 0.383	& 23.09	& 22.36 & disturbed \Ha\\
A370 8792	& 02:39:49.361 	& -01:34:35.081	& 0.369	& 23.25	& 22.58	& disturbed \Ha\\
A370 12154	& 02:39:50.772 	& -01:34:00.510	& 0.368	& 22.49 & 21.82 & disturbed \Ha\\
A370 6295	& 02:39:54.831 	& -01:35:02.081	& 0.360	& 22.91 & 22.29	& faint and small\\

 \hline
    \end{tabular}
    \caption{Other blue galaxies with F606W$<$23.5 and (F606W-F814W$<$0.8). Galaxy ID, coordinates, redshifts and HST magnitudes come from \citealt{Richard+2021}. We give in the last column the spectral classification from the analysis of their MUSE datacubes.
    }
    \label{tab:control}
\end{table*}

Out of these 17 blue galaxies, 12 turned out to be totally passive (k-types), with no sign of emission lines nor of strong Balmer lines in absorption. 8/12 passive galaxies have an HST color in the range
0.7$<$(F606W-F814W)$<$0.8, thus a "green" color still not too far from the red sequence, but other 4 are properly blue with (F606W-F814W)$<$0.7.

Among the remaining five non-RPS, non-PSB blue galaxies with emission lines, one is a clear merger (galaxy A2744 11651), other 3 have a disturbed $\rm H\alpha$ morphology (A370 13663, A370 8792, and A370 12154) in such a way that they are possibly weaker RPS cases (one of them also has unwinding spiral arms (A370 13663), as often found among RPS galaxies at low-z, \citealt{GASPXXIX}), and one is too faint and small to say anything (A370 6295).

To summarize, the population of blue galaxies seen in projection towards the central regions of these two clusters is dominated by two classes of objects: those currently+recently undergoing ram pressure stripping (RPS+PSB), and those that are already completely passive. Together, these two classes account for 85\% of all blue cluster galaxies. The remaining 15\% of the cases, apart from a merger, are only moderately disturbed by some mechanism that cannot be securely identified. 

Excluding the totally passive systems, which are likely to be the end product of RPS+PSB evolution, the fraction of blue galaxies that is seen to be undergoing (RPS) or have recently undergone (PSBs) ram pressure stripping is at least 76\% and becomes 90\% if we include the uncertain weaker cases. This shows that the impact of ram pressure stripping on those galaxies that manage to arrive with still some gas at the center of galaxy clusters at $z \sim 0.4$ is very strong. Ram pressure stripping affects most of (if not all) such galaxies.

\section{The [OII]/\Ha line ratio}\label{sec:o2ha}
The set of gas emission lines that could be observed for these two particular clusters, among those targeted by the MUSE-GTO program, encompasses both the [OII] and the \Ha emission lines, both widely used as star formation indicators.

\begin{figure*}
    \centering
    \includegraphics[width=0.9\textwidth]{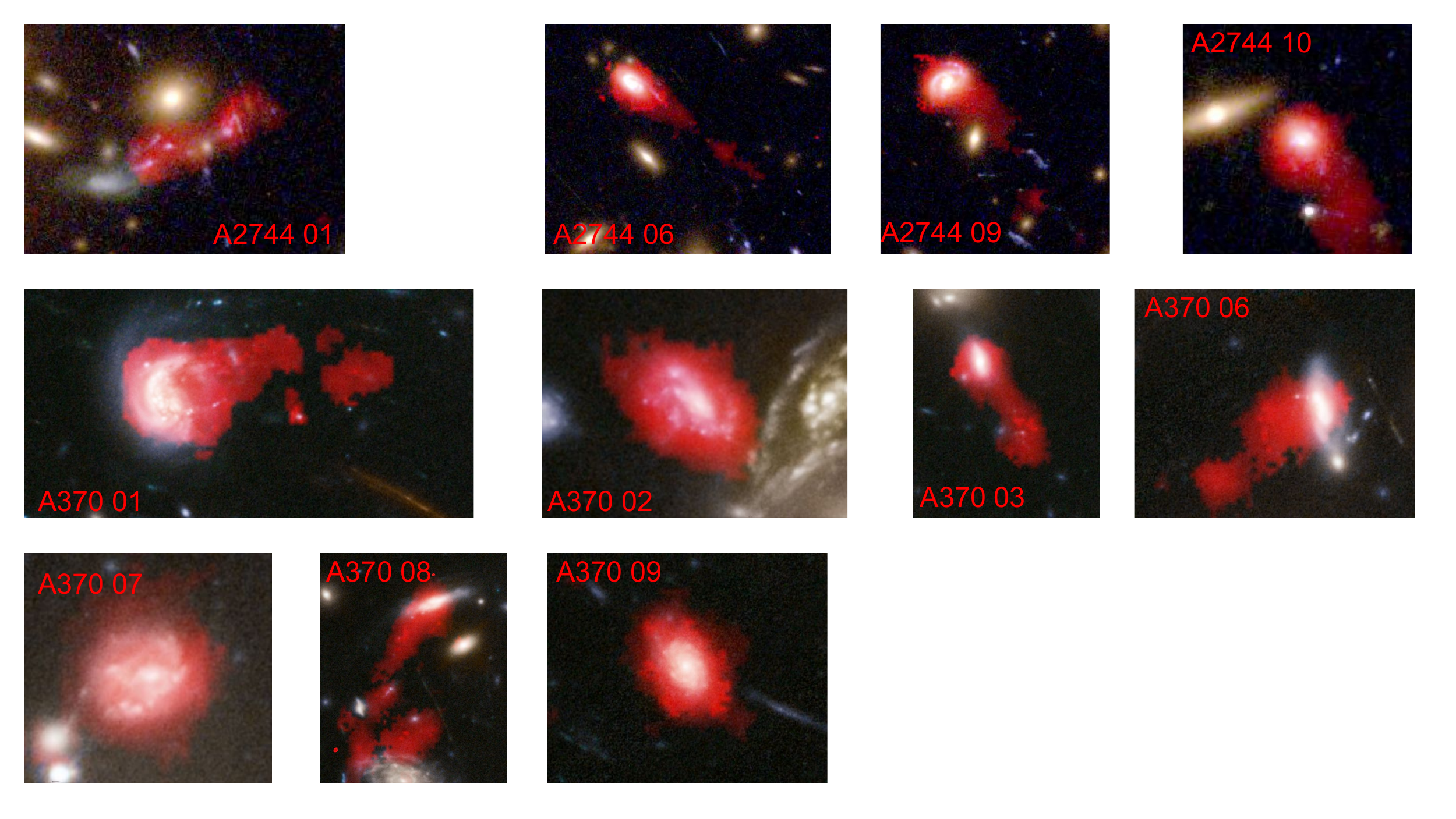}
    \caption{RGB images (from HST) of the RPS galaxies in the analyzed clusters, with superimposed in red the total [OII] emission from MUSE.}
    \label{fig:rps_o2}
\end{figure*}

To better illustrate the extent of the spectacular [OII] tails that characterize RPS galaxies in the inner region of these two clusters, we show in Fig. \ref{fig:rps_o2} the RGB images of the RPS galaxies with superimposed in red the total [OII] emission that we measured from the MUSE datacubes. We show here only the 11 galaxies for which we have been able to model and subtract the stellar component (therefore excluding A2744 03 and A2744 04).

As shown in Fig. \ref{fig:rps_o2}, we found that the [OII] emission is very prominent in the tail, and it is worth comparing it with the corresponding \Ha emission.

To our knowledge, this is the first time an [OII]/\Ha ratio can be spatially resolved  within the galaxy disk and in the tails of ionized gas of jellyfish galaxies.
The only measurement available so far of the [OII]/\Ha ratio in the stripped gas of local galaxies is the one given in \citet{Cortese+2006}, who found [OII]/\Ha$\sim$2 in the tail of possibly pre-processed galaxies infalling into the nearby cluster A1367, using MOS/longslit spectroscopic observations. [OII] and \Ha have also been measured in one star forming knot in the gas stripped tail of 235144-260358 in A2667 at redshift $\sim$ 0.2 \citep[][]{Cortese+2007}, with \Ha line dominating over the [OII].

\begin{figure*}
    \centering
    \includegraphics[width=0.9\textwidth]{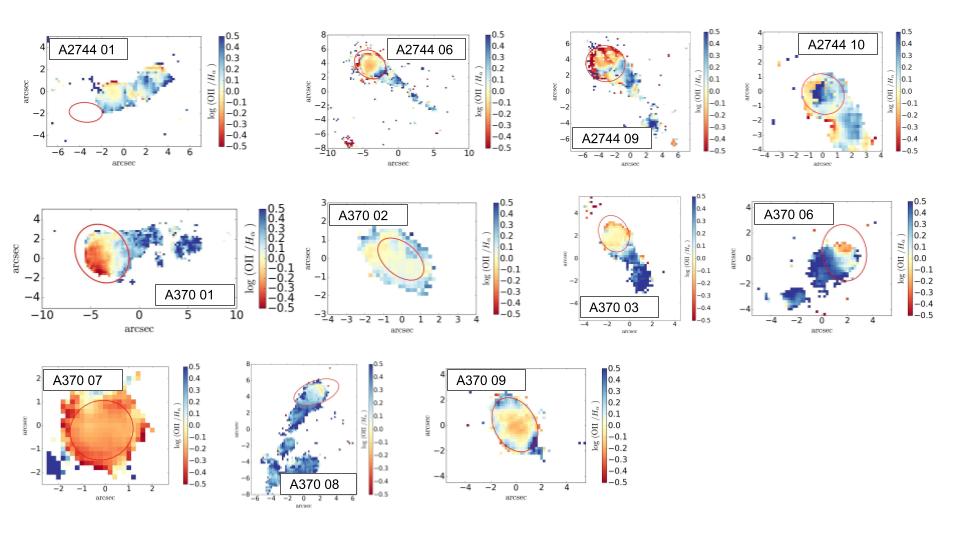}  
    
    \caption{[OII] over \Ha flux ratio for spaxels where both lines have been detected with SN$>2$: red regions are dominated by \Ha, blue regions by the [OII]: galaxies tails are [OII] dominated. The red ellipses trace the stellar disk. Line fluxes are measured on the emission only datacube and dust corrected.}
    \label{fig:o2_ha}
\end{figure*}

Fig. \ref{fig:o2_ha} shows the [OII]/\Ha dust corrected flux ratio maps for the ram pressure stripped galaxies in A2744 and A370, with red colors representing \Ha dominated spaxels ([OII]/\Ha$<1$) and blue colors [OII] dominated spaxels ([OII]/\Ha$>1$).

The average [OII]/\Ha values within the stellar disk are in agreement with the ones found in 
low-z star-forming galaxies in
the NGFS \citep{Kewley+2004} and in the 2dFGRS \citep{Mouhcine+2005}. In several of our RPS galaxies, (A2744 06, A2744 09, A370 01, A370 02, A370 03, A370 06, A370 08) 
the inner part of the stellar disk is  dominated by the \Ha emission, with the lowest [OII]/\Ha emission shown at the compression front generated by the galaxy motion within the ICM.

At the opposite side of the compression front, i.e. where the gas tail has developed, the [OII]/\Ha ratio is completely inverted: tails are dominated by the [OII] emission, which starts to be dominant close to the stellar disk limit that we have defined on the basis of the stellar emission.
The values that we find in all the ram-pressure stripped tails are in agreement with the ones obtained by \cite{Cortese+2006}, ranging from 1 to 3.

The peculiar [OII]/\Ha line ratios can also be used to interpret the emission in the galaxies that do not show spectacular tails: A370 07 and A370 09 show a similar behaviour (high [OII]/\Ha ratios outside of the stellar disk)
 but their tail is probably hidden behind the main galaxy body.
The high values we find for the [OII]/\Ha ratio in the ram-pressure stripped tails could 
be due to either a global decrease in metallicity \citep[][]{Mouhcine+2005}, which does not seem to be the case in our jellyfish tails because there gas metallicities are high (paper in prep.), and/or the lower gas densities that characterize these regions, as in this case the oxygen can not be efficiently collisionally de-excited. 
Such high values of [OII] have been found, for instance, in the diffuse regions of starburst galaxies \citep{Martin1997}.
We note also that in the nearby GASP ram-pressure stripped galaxies the stripped tails are dominated by the Diffuse Ionized Gas (DIG) emission \citep[][]{GASPXXXII, GASPXXXV}, possibly linked to a less dense gas (showing lower \Ha emission and Composite/LINER like features).

 However, anomalous emission-line ratios might also be due to the interaction with the hot plasma surrounding the stripped ISM (either cooling of the ICM or mixing ISM-ICM), as suggested also by X-ray results \citep[][]{GASPXXIII, GASPXXXIV}.
Only a careful analysis of all the involved line ratios that can now be observed with MUSE will shed light on the ionization mechanisms and gas conditions in the stripped tails. 

\section{Conclusions}\label{sec:conclusions}
In this paper we have presented the characterization of 13 RPS galaxies located in the central region of the two intermediate redshift ($\sim0.3$) clusters A2744 and A370. Six more galaxies (PSBs) appear to have been affected by the same physical mechanism and will be discussed in a separate paper (Werle+, in prep.).
To infer the presence of the ram pressure stripping we have made extensive use of the MUSE-GTO data, which, despite covering only the inner cluster regions, have been fundamental to characterize the galaxies and their stripped tails, while HST data can only suggest the presence of ram-pressure stripped candidates. 
At low redshift, ram pressure stripped galaxies with tails have been observed over a wide range of clustercentric distances, up to the cluster virial radius \citep[][]{Yagi2017,Gavazzi+2018,Roberts+2021}, but it is close to the cluster center that the most striking {\sl ionized gas tails} have been found \citep[][]{GASPIX,GASPXXI}. If the same happens at intermediate redshifts, the MUSE data of GTO clusters may been giving us a view of many of the most spectacular cases of stripped tails of ionized gas.

Galaxies showing the presence of ram-pressure stripped tails have stellar masses from $\sim 10^8$ to $\sim 10^{10.5}$ \Msun.
Ram-pressure stripped galaxies at redshift $0.3-0.4$ are characterized by long (7 to $\sim 50$ kpc) shrinking tails, with projected line of sight velocities that tend to increase moving farther from the galaxy disks.
Their ionization is powered by star formation (in the disks), according to the BPT diagram, that shows line ratios typical of photoionization models, or extending toward the so-called composite region in the tails.
The possibility to extend the observed wavelength range covering the [OII] doublet has allowed us to infer for the first time the spatially resolved  [OII]/\Ha line ratio, finding that galaxy tails are dominated by the [OII] emission, while the opposite is true in galaxy disks.

The great majority of blue cluster members in the central regions of A2744 and A370 are either clearly 
undergoing RPS with tails, or they display a completely quenched star formation as implied by the lack of \Ha emission, being PSBs or totally passive with weak Balmer lines in absorption.

These results show beyond doubt that ram -pressure stripping is at work in distant clusters and it is the dominant mechanism for depriving galaxies of their gas in the central regions of clusters up to at least z$\sim$0.4.

\acknowledgments
{We wish to thank the anonymous Referee for the very constructive report, that helped us improving the paper. This project has received funding from the European Research Council (ERC) under the European Union's Horizon 2020 research and innovation programme (grant agreement No. 833824, GASP project). 
We acknowledge funding from the agreement ASI-INAF n.2017-14-H.0 (PI A. Moretti), as well as from the INAF main-stream funding programme.
B.~V. and M.~G. also acknowledge the Italian PRIN-Miur 2017 (PI A. Cimatti).
We acknowledge Y. Jaff\'e for the stimulating discussion.
This research made use of APLpy, an open-source plotting package for Python (Robitaille and Bressert, 2012; Robitaille, 2019).
}

\software{pPXF, SINOPSIS, HIGHELF, KUBEVIZ
}
\newpage
\appendix \label{sec:appendix}
In the following for every RPS galaxy we present a series of panels that illustrate the ram pressure stripping effects.

In each plot the black ellipse shows the extent of the stellar disk, derived as explained in Sec. \ref{sec:data_analysis}, while the black circle (generally in the lower right corner) shows the size of the typical PSF of the MUSE images that we used to smooth the data (i.e. 1 arcsec).
The colored spaxels in all the plots have been selected to have a global S/N (calculated over the entire zero-moment of the cube) larger than 2, and for which we have been able to measure both the velocity and the velocity dispersion of the reference line (here \Ha or [OII]). We excluded spaxels where the velocity or the velocity dispersion was more than 3$\sigma$ deviant from the median distribution.
\Ha and [OII] fluxes have been measured on the emission only datacubes, i.e. those obtained after having subtracted the stellar contribution, and have then been corrected for internal extinction according to the measured Balmer decrement.

More in detail, next figures contain the \Ha and [OII] fluxes in erg/cm$^2$/s/arcsec in logarithmic units (top left and top right of the first row, respectively); the \Ha line of sight velocity and the stellar velocity in \kms (left and right panels in the second row); the \Ha and stellar velocity dispersions in \kms (left and right panels in the third row);
the A$_V$ extinction derived from the Balmer decrement and the map of the galaxy where each spaxel is color-coded according to its classification in the BPT diagram \citep[][]{BPT} involving the [NII] line, as done in GASP \citep[][]{poggianti2017}. Red spaxels are those dominated by star formation, orange ones are classified composite, blue spaxels lie in the LINER/shocked regions and green ones are characterized by line ratios typical of AGN.
Each plot also shows a scale corresponding to 10 kpc at the cluster redshift, assuming a (70, 0.3, 0.7) cosmology.
Given the limited spatial resolution of MUSE ($\sim$1\arcsec), we expect that the effect of beam smearing is quite strong, in particular for poorly resolved galaxies. Therefore, it is not possible to disentangle between the contribution of the rotation velocity and the velocity dispersion to the broadening of emission lines by simply looking at the velocity field. Moreover, the spectral resolution of MUSE observations ($\sim$50 km/s) further limits our ability of disentangling the two contributions.

\begin{figure*}
    \centering
    \includegraphics[width=0.45\textwidth]{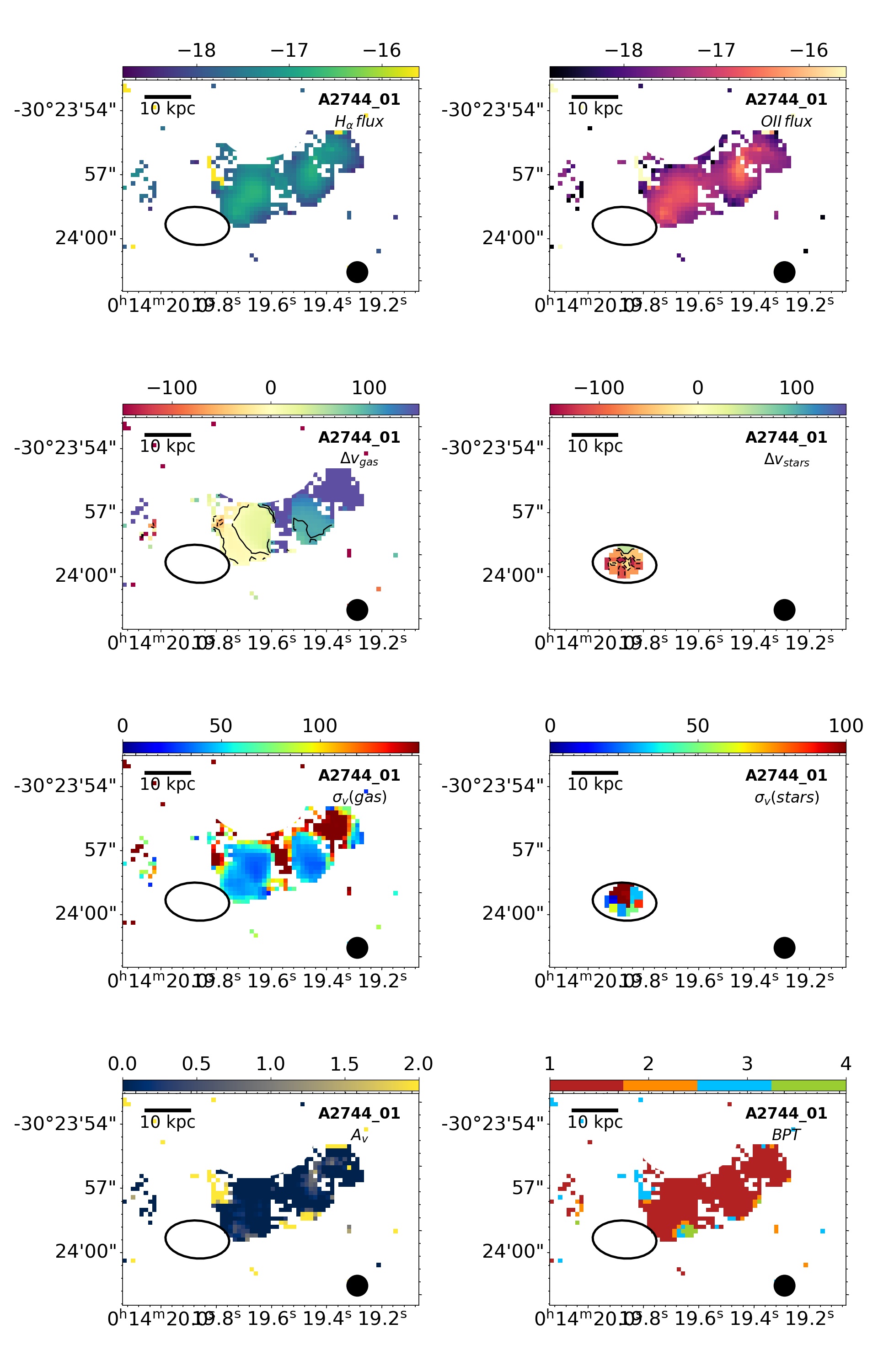}
   \includegraphics[width=0.45\textwidth]{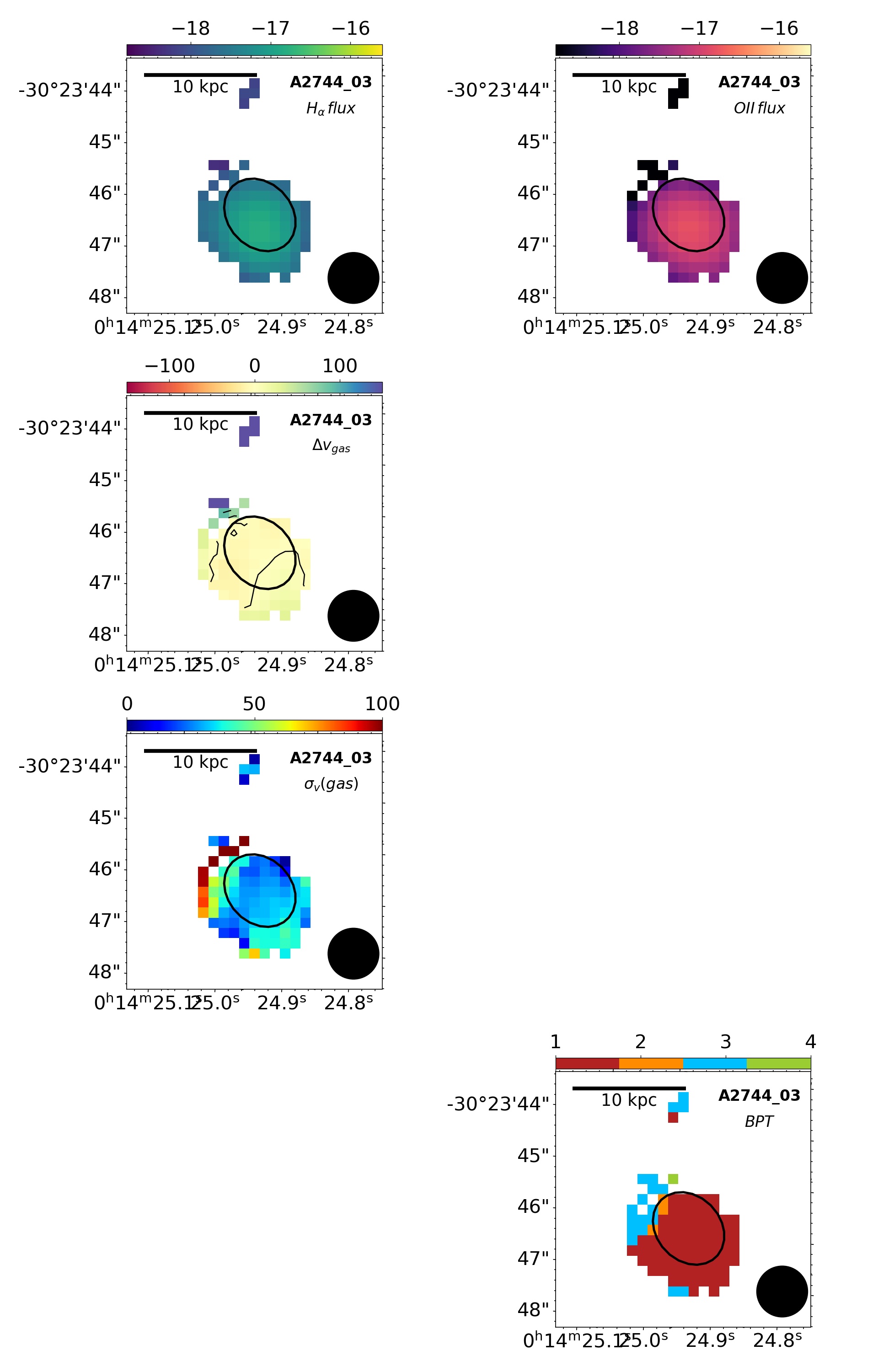}
    \caption{\Ha and [OII] fluxes, \Ha and stellar velocities, \Ha and stellar velocity dispersions, A$_v$ and BPT map for A2744 01 (two left columns) and A2744 03 (two right columns). Stellar kinematics could not be derived for A2744 03 due to the low S/N.}
    \label{fig:a2744_01_03}
\end{figure*}

\begin{figure*}
    \centering
    \includegraphics[width=0.45\textwidth]{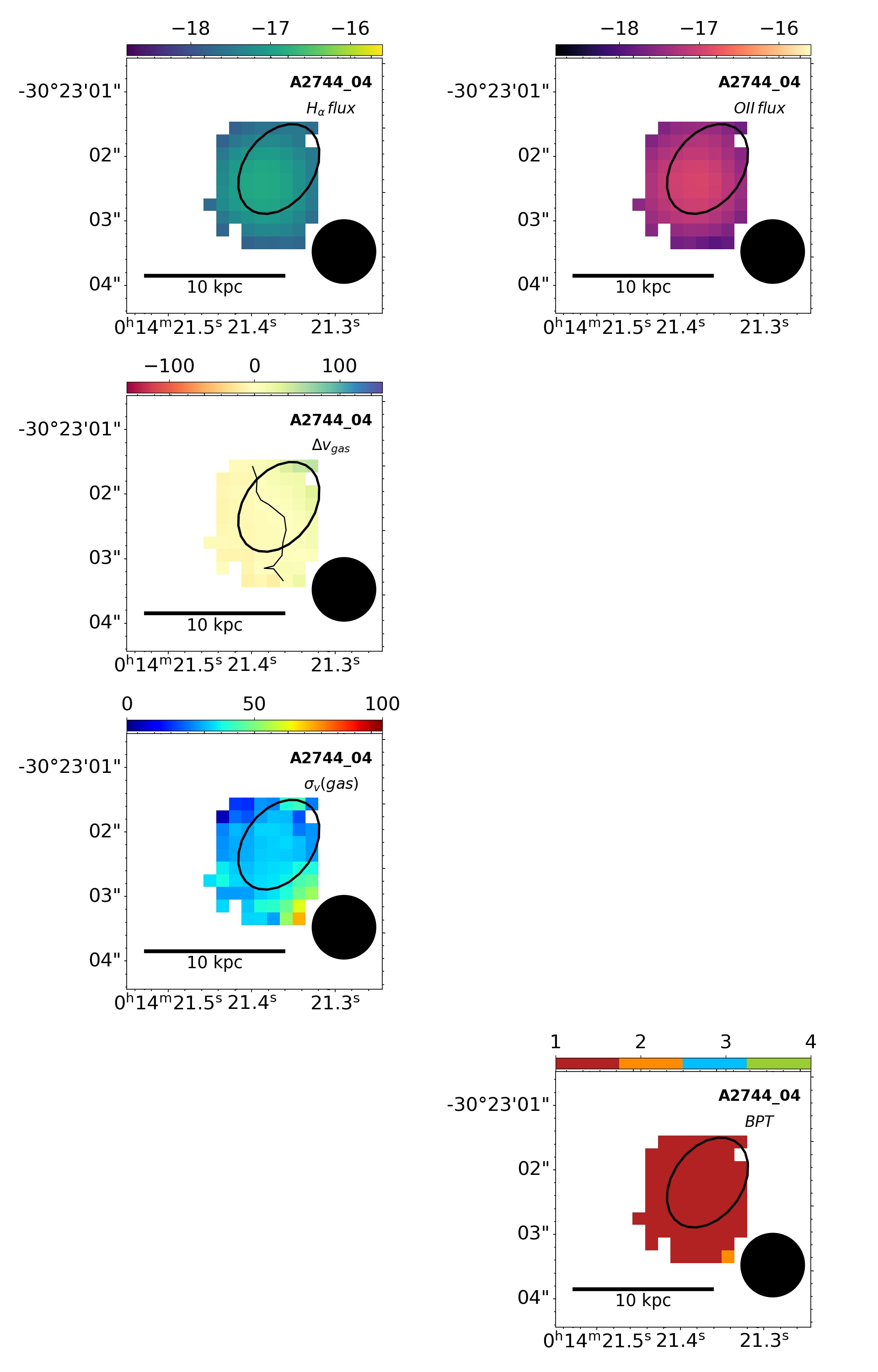}
   \includegraphics[width=0.45\textwidth]{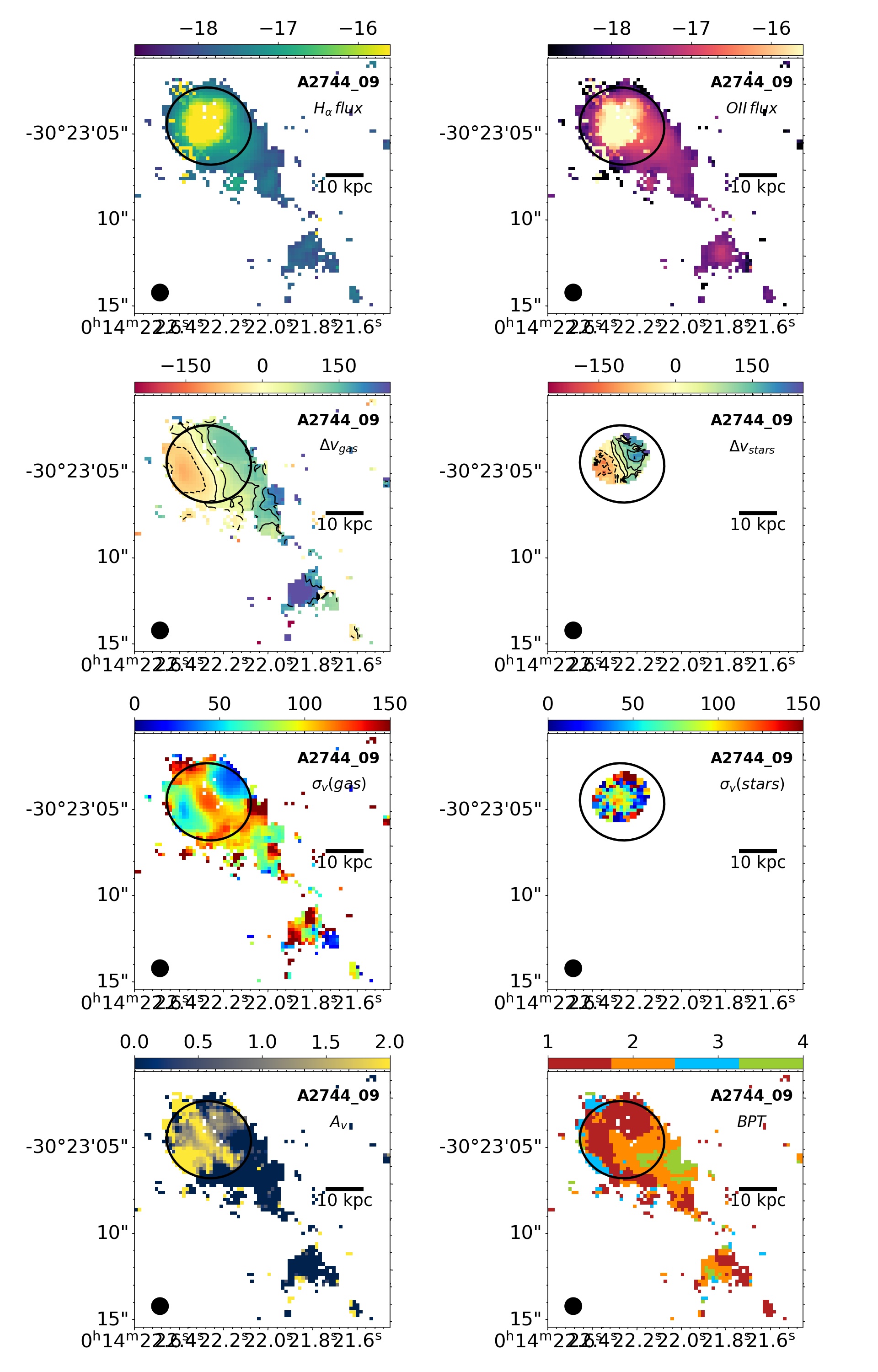}
    \caption{\Ha and [OII] fluxes, \Ha and stellar velocities, \Ha and stellar velocity dispersions, A$_v$ and BPT map for A2744 04 (two left columns) and A2744 09 (two right columns). Stellar kinematics could not be derived for A2744 04 due to the low S/N.}
    \label{fig:a2744_04_09}
\end{figure*}

\begin{figure*}
    \centering
    \includegraphics[width=0.45\textwidth]{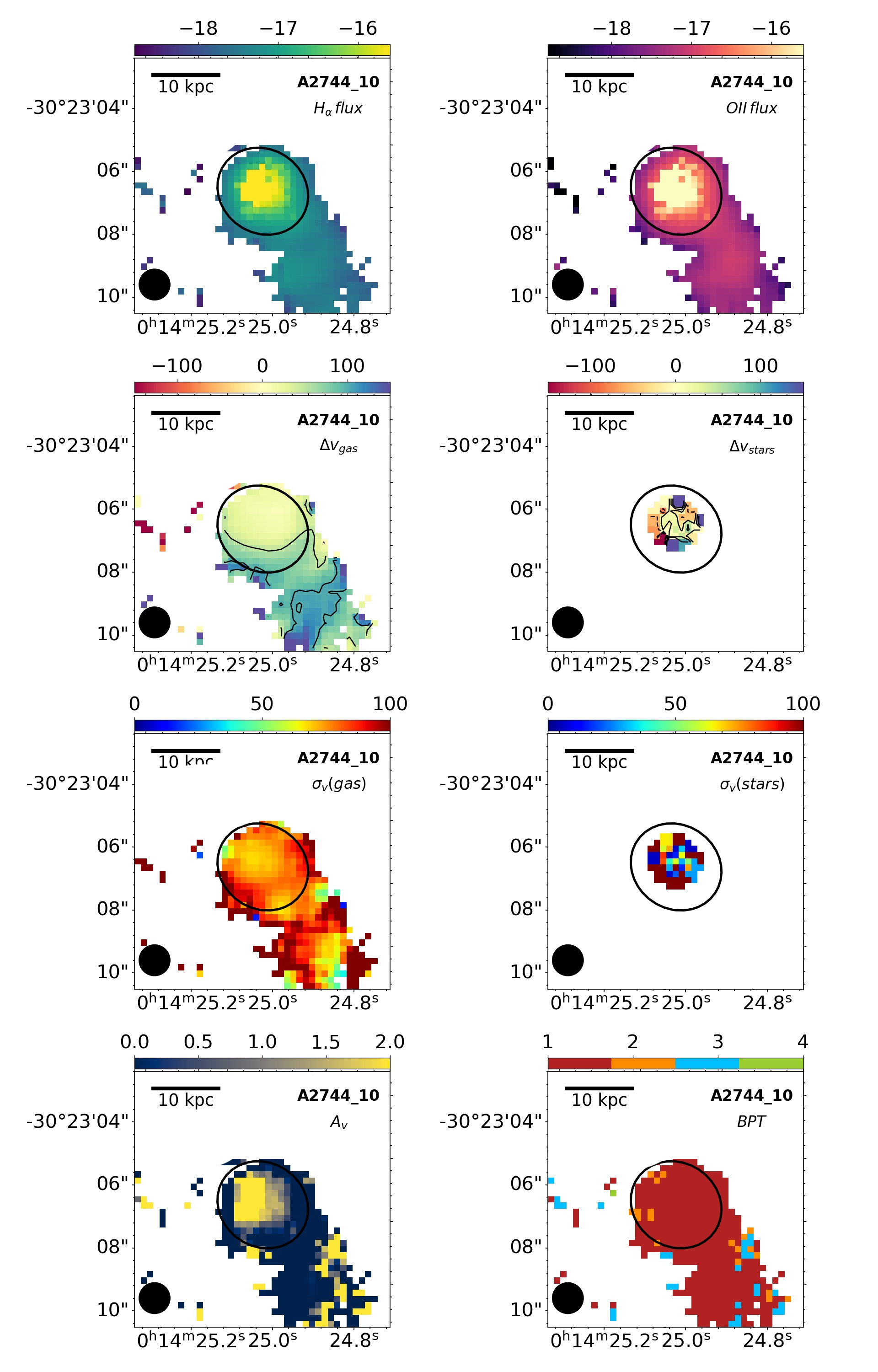}
    \includegraphics[width=0.45\textwidth]{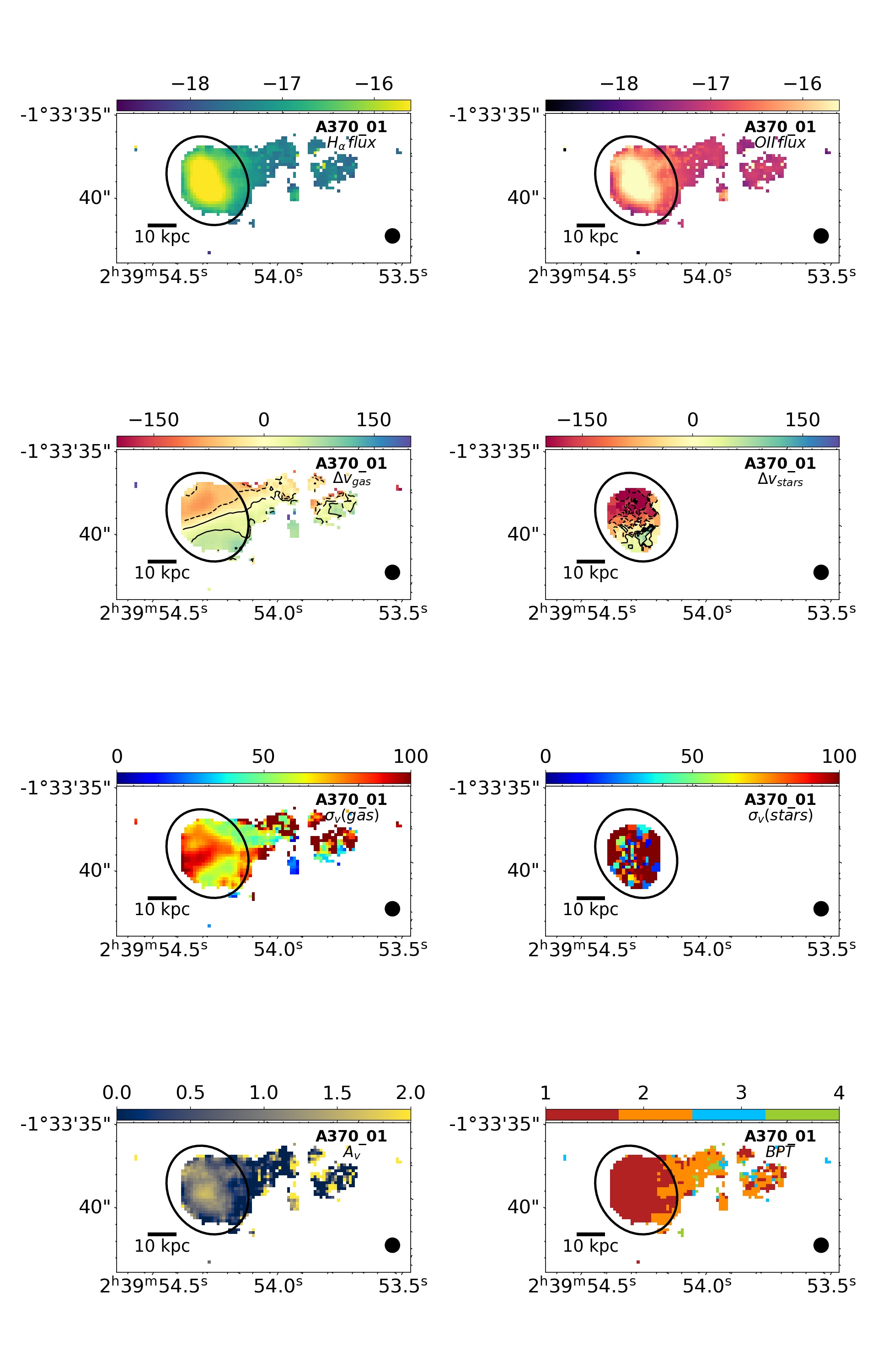}
    \caption{\Ha and [OII] fluxes, \Ha and stellar velocities, \Ha and stellar velocity dispersions, A$_v$ and BPT map for A2744 10 (two left columns) and A370 01 (two right columns).}
    \label{fig:a370_01_02}
\end{figure*}

\begin{figure*}
    \centering
   \includegraphics[width=0.45\textwidth]{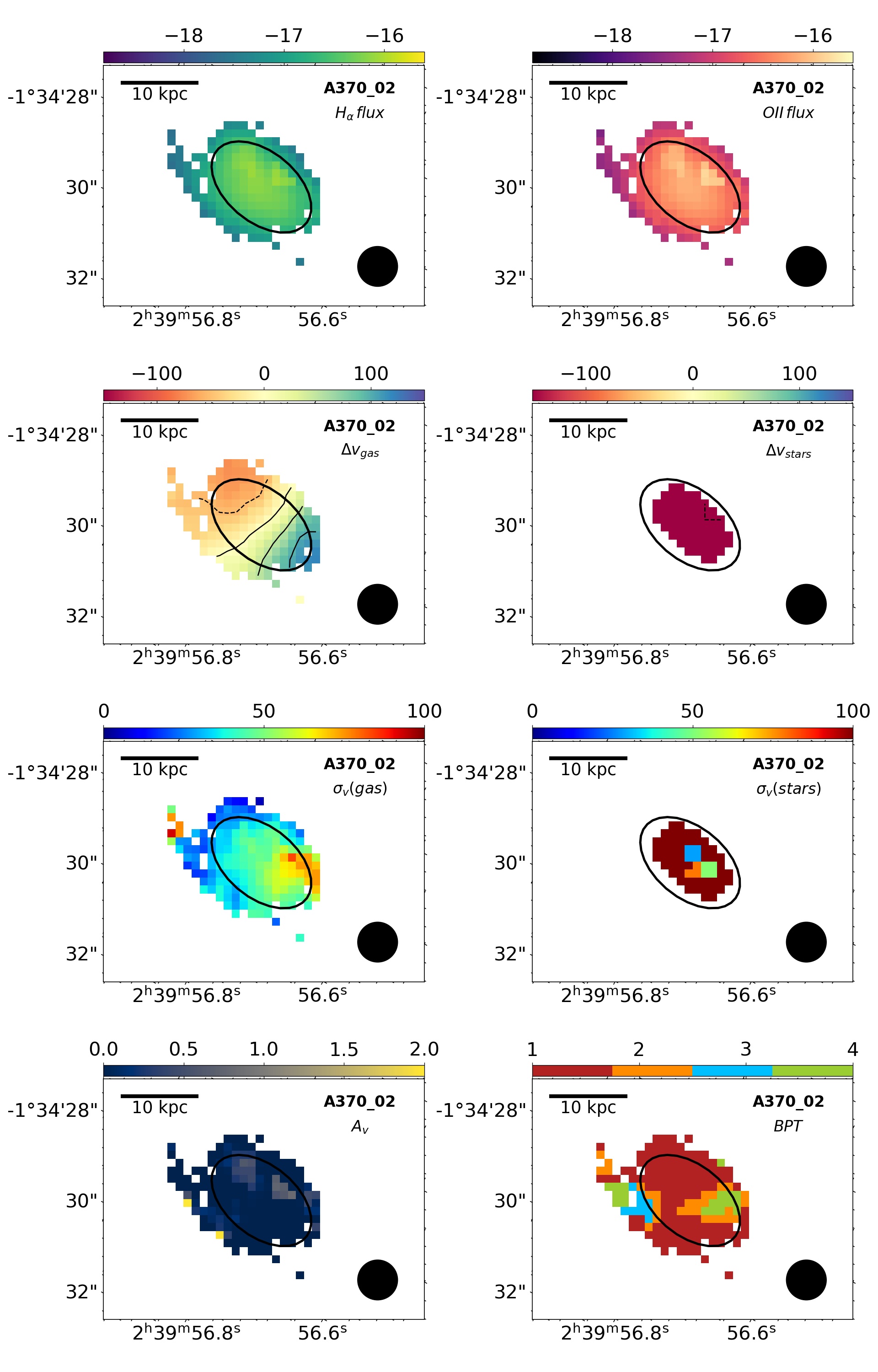}
    \includegraphics[width=0.45\textwidth]{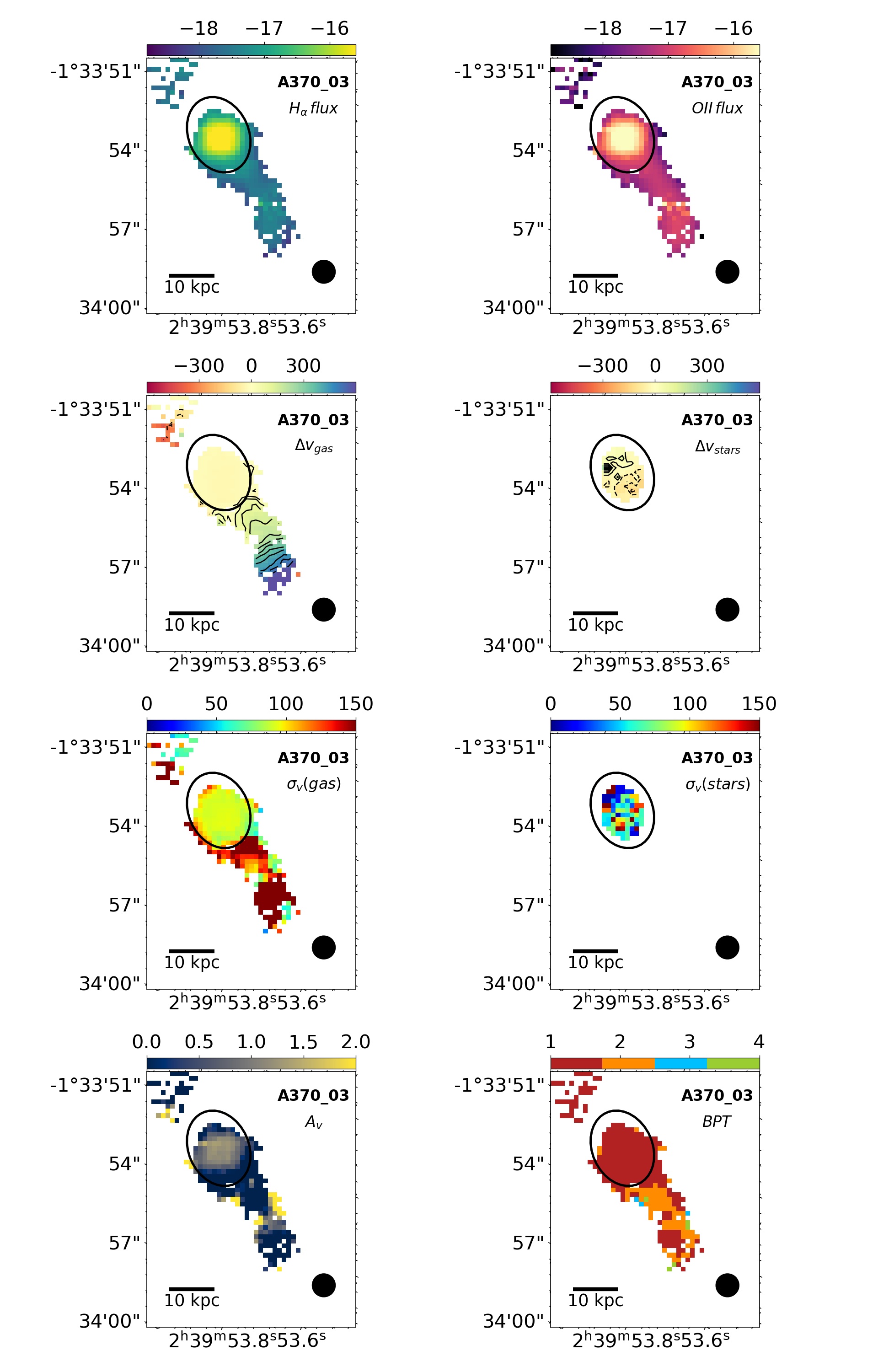}
    \caption{\Ha and [OII] fluxes, \Ha and stellar velocities, \Ha and stellar velocity dispersions, A$_v$ and BPT map for A370 02 (two left columns) and A370 03 (two right columns).}
    \label{fig:a370_03_06}
\end{figure*}

\begin{figure*}
    \centering
   \includegraphics[width=0.45\textwidth]{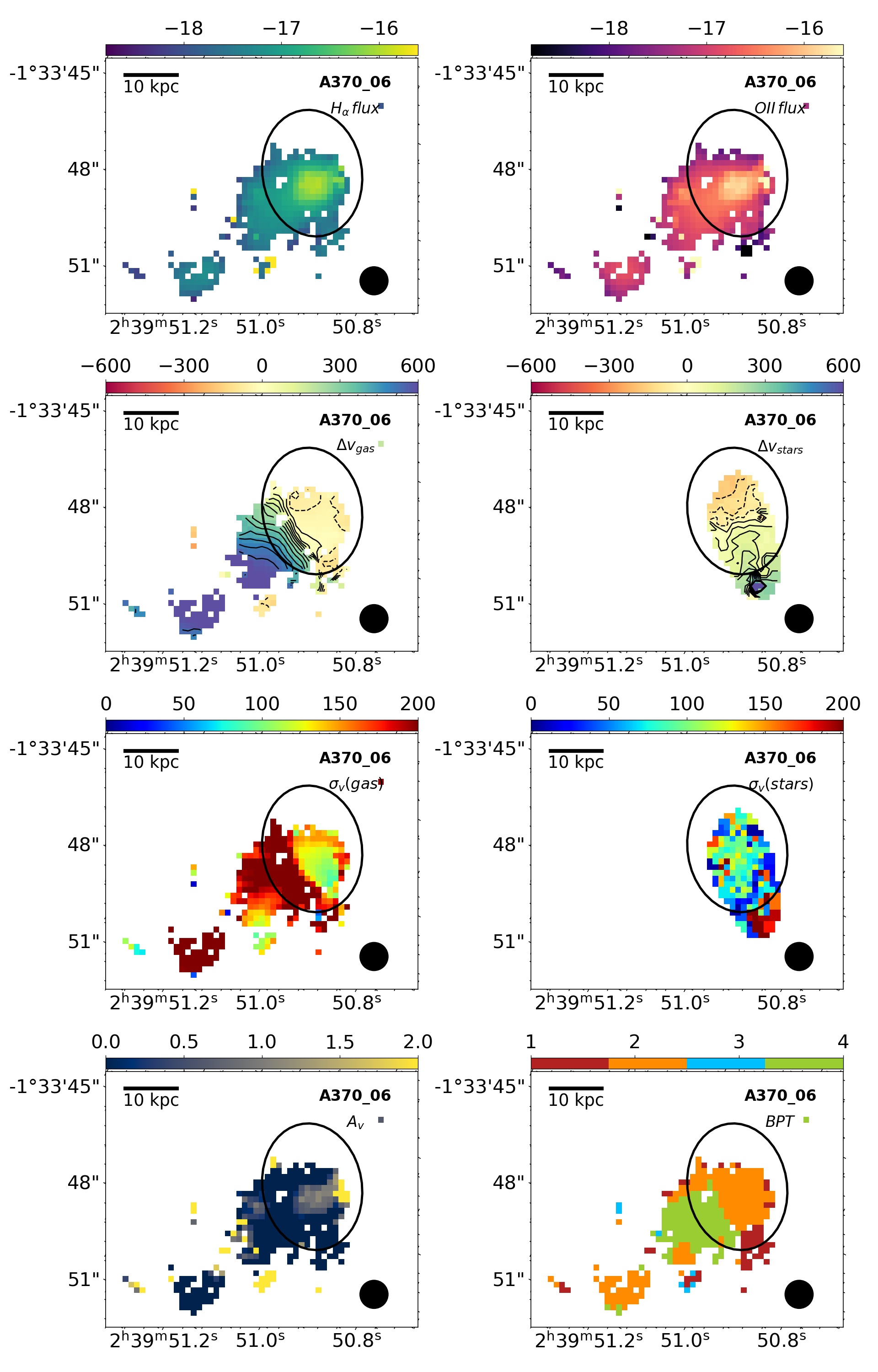}
    \includegraphics[width=0.45\textwidth]{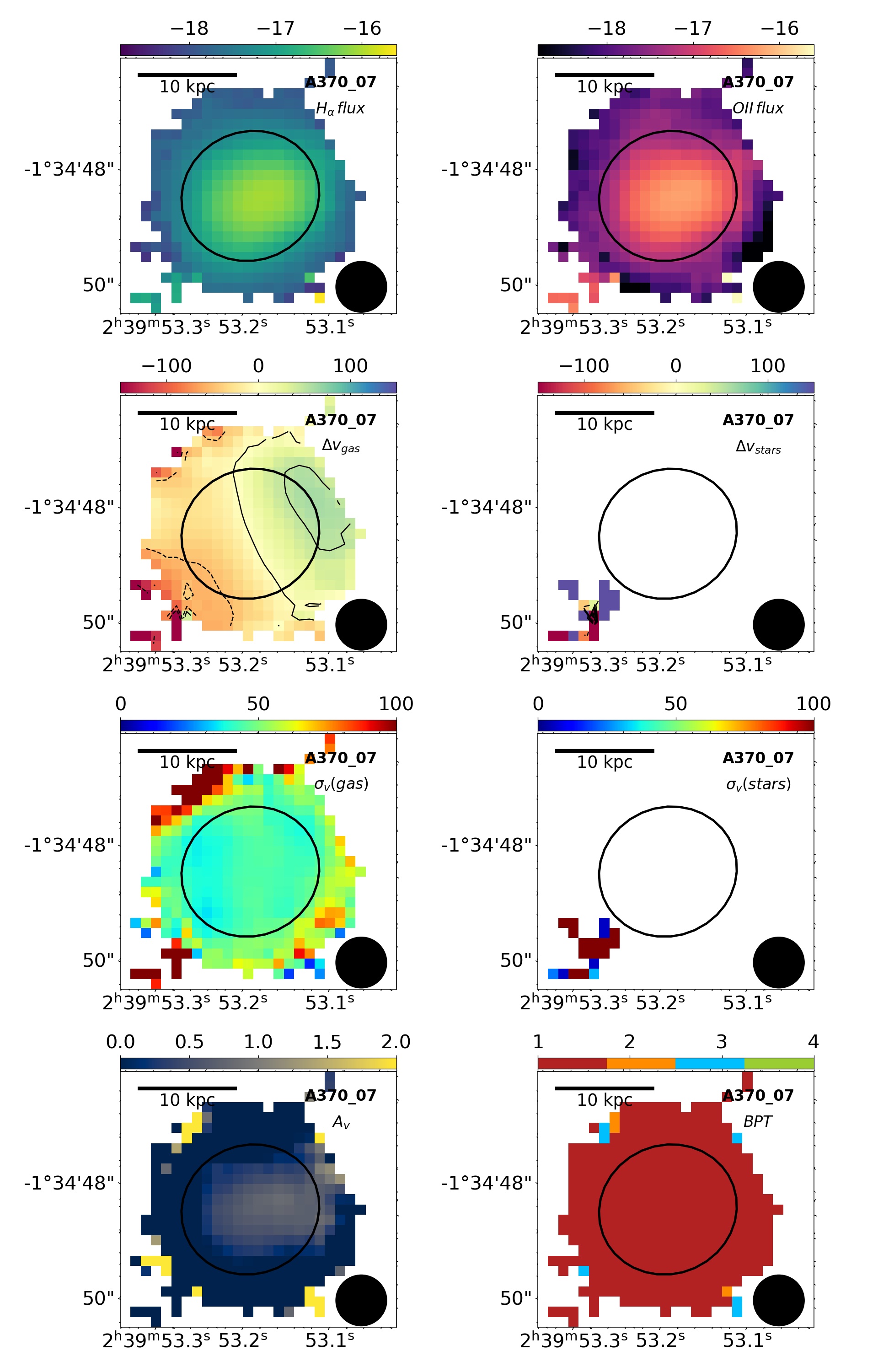}
    \caption{\Ha and [OII] fluxes, \Ha and stellar velocities, \Ha and stellar velocity dispersions, A$_v$ and BPT map for A370 06 (two left columns) and A370 07 (two right columns).}
    \label{fig:a370_07_08}
\end{figure*}

\begin{figure*}
    \centering
   \includegraphics[width=0.45\textwidth]{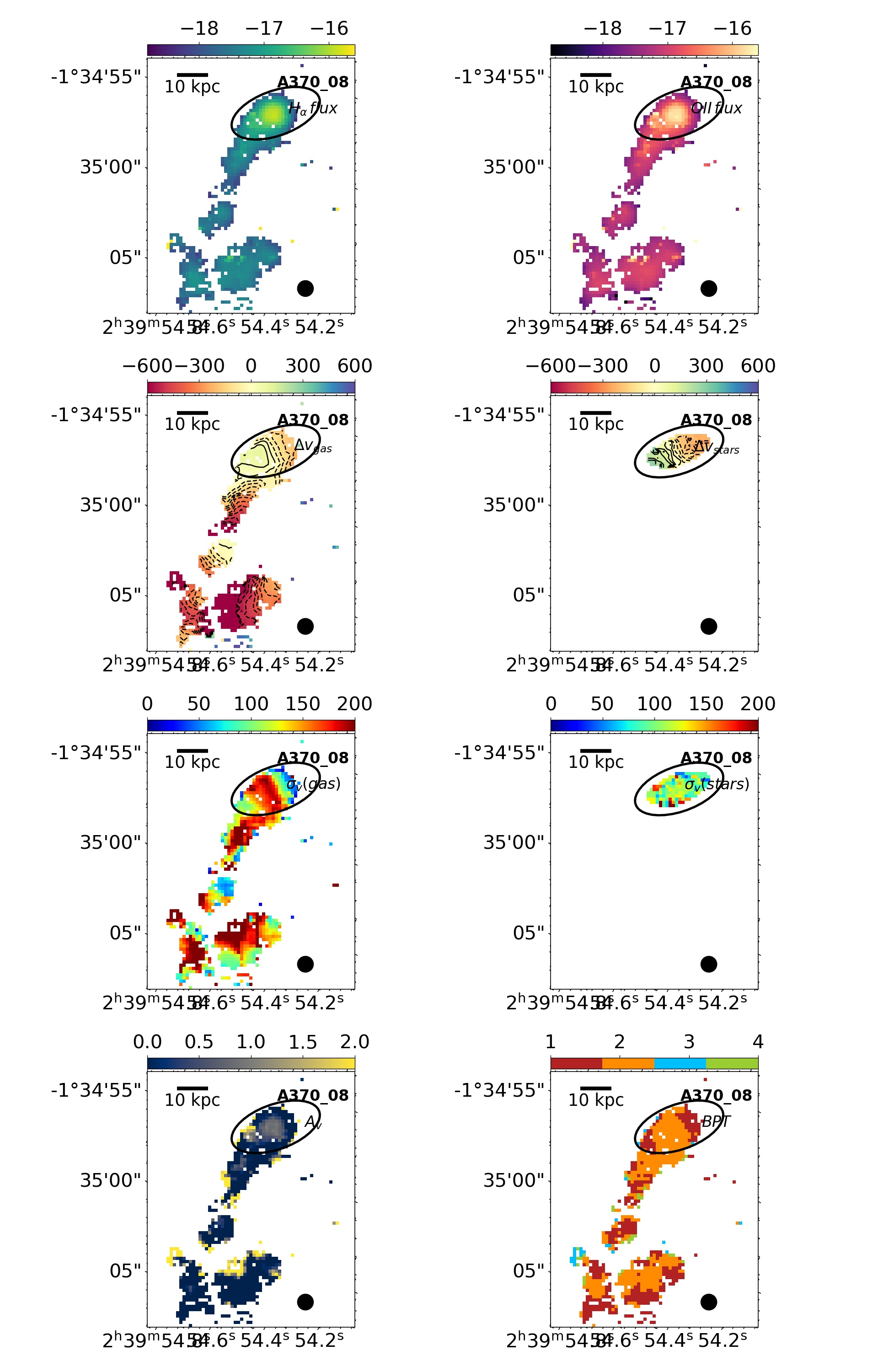}
    \includegraphics[width=0.45\textwidth]{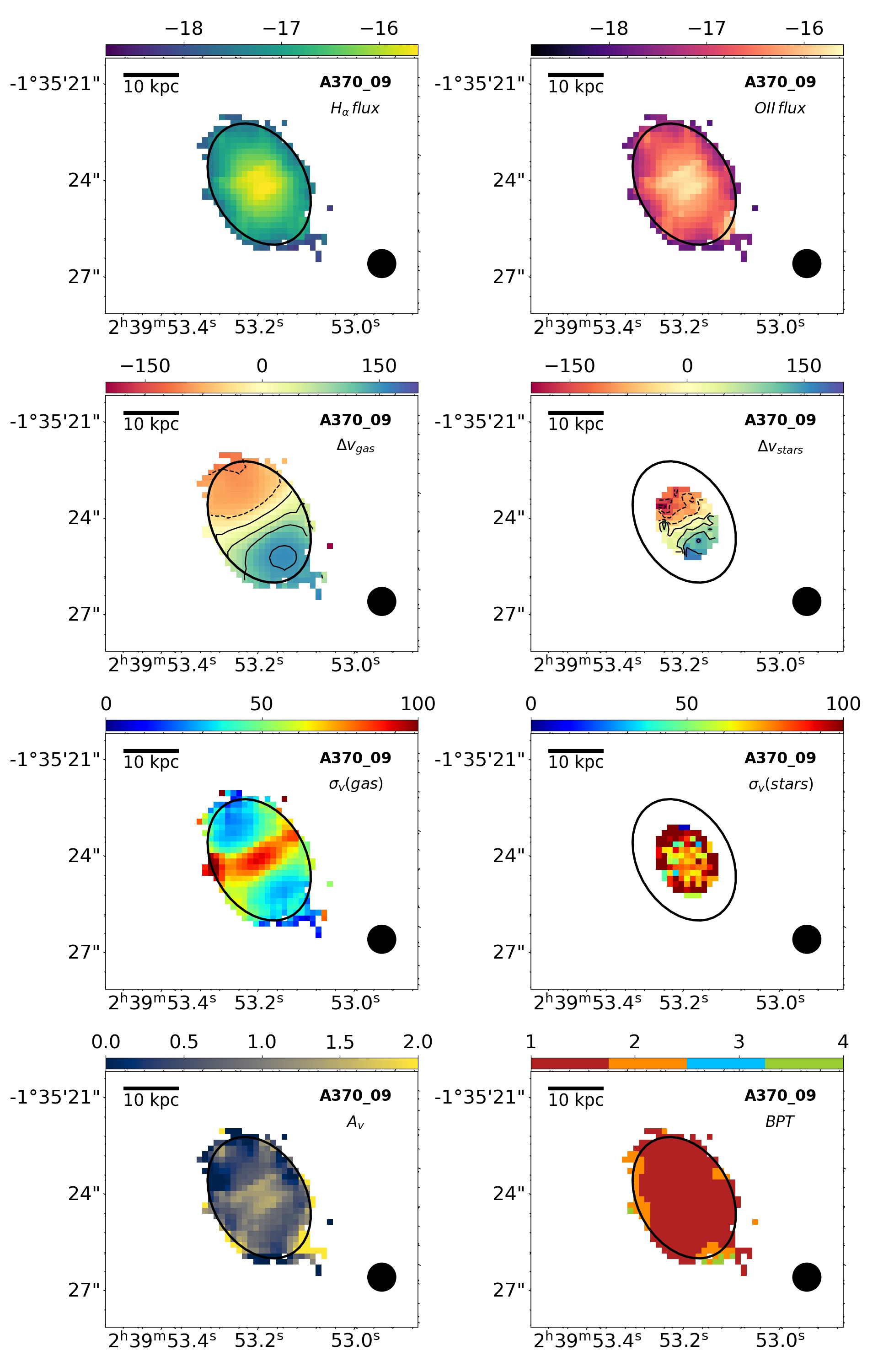}

    \caption{\Ha and [OII] fluxes, \Ha and stellar velocities, \Ha and stellar velocity dispersions, A$_v$ and BPT map for for A370 08 (two left columns) and A370 09 (two right columns).}
    \label{fig:a370_09}
\end{figure*}
\begin{itemize}

    \item A2744 01.
    In the region shown in the upper left panel of Fig. \ref{fig:RGB} there are multiple galaxies, with different colours. In particular a long tail of blue knots  extending from east to west is clearly seen south of the brighter red galaxy. Both this galaxy and the small red object in the middle of the blue knotty tail are cluster members, with a redshift $\sim 0.30$. 
From the analysis of the emission line redshift we deduce that the parent galaxy of the blue emitting knots is instead the one enclosed within the red contour in Fig. \ref{fig:RGB} that traces the 3 sigma contour above the background emission.
This galaxy was already classified as jellyfish by \cite{Owers+2012}, but no information were available about its redshift.
Thanks to the MUSE data, we can now infer that this galaxy is a cluster member, with a stellar mass of $10^{8.5}$ \Msun and a redshift of $\sim 0.2913$.
The tail of stripped gas that is still star forming extends out to 8 arcsec from the galaxy center, i.e. $\sim 35$ kpc, while the galaxy itself is completely devoid of gas (a classical PSB galaxy), see Fig.~\ref{fig:a2744_01_03}. The tail is visible both in \Ha and [OII] (first and second panels in the first row of Fig.\ref{fig:a2744_01_03}, respectively).
The gas velocity with respect to the main body, as derived from the \Ha redshift, is increasing from 0 to $\sim 200$ \kms at the largest distance (left panel in the second row of Fig.\ref{fig:a2744_01_03}), confirming that this galaxy is being stripped of its gas while moving along the line of sight toward us in the cluster space.
The region at $\sim 300$ \kms at the west end of the tail seems to be associated with another small galaxy, as the analysis of the gas velocity dispersion confirms (left panel in the third row of Fig.\ref{fig:a2744_01_03}).
The stellar kinematics is dominated by the beam smearing effect (the typical PSF size of 1 \arcsec is shown as a black circle in the bottom right corner of each plot).
The extinction is everywhere low ({typically \bf $A_V<0.5$)}, and the analysis of the line ratios indicates that the entire gaseous tail is ionized by star formation (last row of Fig.\ref{fig:a2744_01_03}).
HST images confirm the presence of blue star forming regions in correspondence of the \Ha emission, as visible in Fig. \ref{fig:RGB}.

\item 
A2744 03 is a very low mass galaxy ($\log(M_{\star}/M_{\odot})=7.98$), probably a dwarf irregular, for which we could not measure the stellar kinematics.
It is clearly star forming, and it has a disturbed morphology, reminiscent of a tail pointing toward the north of the image and in the HST RGB image (Fig. \ref{fig:RGB}), even if the ionized gas extends to the south of the stellar disk.
There is no appreciable stellar continuum, but the kinematics of the gas shows a hint of a small rotation ($\sim 30$\kms), entirely similar to the gaseous velocity dispersion.
The small size of this galaxy ($\sim 1.5$ arcsec) is very similar to the PSF of the MUSE images, and we could actually be dominated by the beam smearing.
No other possible parent galaxy is seen close to this object.
We show in Fig.\ref{fig:a2744_01_03} both the \Ha velocity and its velocity dispersion, measured on the original MUSE datacube.
We did not calculate the dust extinction from the emission lines, as we could not subtract the stellar continuum. 
The BPT map has been derived therefore on the emission lines not corrected for internal dust.
The entire galaxy seems star forming, while the tail appears to be LINER-like.

\item 2744 04
As the previous galaxy, also A2744 04 is a low surface brightness dwarf galaxy, with a stellar mass of ($\log(M_{\star}/M_{\odot})= 8.18$). It shows a tail extending toward the north-west (see Fig. \ref{fig:RGB}) direction and again it shows no detectable stellar continuum. No other galaxies are present in the surroundings, and we therefore think of this galaxy as a small star forming galaxy with a tail of stripped gas.
The \Ha emission seems to show a small rotation ($\sim 50$\kms), again compatible with the gas velocity dispersion. The small size of this galaxy suggests that we can not disentangle the two contributions (velocity dispersion and rotation velocity), though.

\item 
A2744 09 is one of the  most massive galaxies subject to ram-pressure stripping in A2744 ($\log(M_{\star}/M_{\odot})=10.57$).
From the main body of the galaxy a long gas tail extends towards south-west, partially maintaining the velocity gradient of the gas in the disk, at least in the first part of the tail.
Again, the stellar kinematics is undisturbed and substantially coincident with the gas rotation in the disk ($\sim 150$ \kms).
The small galaxy south from the main body of A2744 09 at redshift $\sim 0.315$ seems to be not interfering with the stripped gas, nor disturbing the disk gas kinematics in an appreciable way. 
The tail length is $\sim 10$ arcsec i.e. $\sim 42$ kpc at the cluster redshift.
The gas velocity dispersion has a maximum at the galaxy center of $\sim 150$ \kms while the central stellar velocity dispersion is slightly lower. 
The gas extinction is very low in the tail, but its value rises within the stellar disk in the east part where the line ratios indicate the presence of a LINER-like, possibly shocked region (two lowest panels in the right columns of Fig. \ref{fig:a2744_04_09}). The galaxy disk is dominated by star formation extending towards the composite region in the BPT diagram. Small zones of AGN-like emission are found in the initial part and at the extremes of the stripped tail, maybe reminiscent of an old ionization cone.

\item
A2744 10 is a rather low-mass galaxy ($\log(M_{\star}/M_{\odot})=9.3$) with a pronounced tail in the south-west direction extending for $\sim$ 13 kpc and clearly visible both in \Ha and in the [OII]. Within the stellar disk, it shows a low gas rotation ($\sim$ 50 \kms) and velocity dispersion ($\sim$ 70 \kms), while the stellar kinematics is much more confused, due to the small size of this galaxy. The gas velocity increases along the tail extension up to $\sim$ 150 \kms. The A$_V$ is high at the galaxy center, and decreases to low values in the external zone of the disk and in the tail, with a decrease in the outer part of it, in correspondence with a region with lower projected velocity and some spaxels apparently dominated by LI(N)er--like emission. From the analysis of the emission lines involved in the BPT diagram it turns out that the whole disk and the stripped tail are dominated by star formation. The [OII]/\Ha map shown in Fig. \ref{fig:o2_ha} reveals within the disk a bimodal behaviour: half of the disk seem dominated by the \Ha emission (the one at the compression front) and a second half dominated by the [OII], that is also predominant in the tail.

\item 
A370 01 is the most massive galaxy in the A370 sample ($\log(M_{\star}/M_{\odot})=10.62$) and shows a ionized gas tail $\sim 8$ arcsec long, i.e. $\sim 38$ kpc (first and second panels in the first row of Fig.\ref{fig:a370_01_02} for the \Ha and [OII] emission, respectively). Part of the tail is visible from the HST images, but is outside the MUSE observed field. As in A2744 09 the gas rotation pattern is retained along the tail, that departs from a bright spiral arm in the galaxy.
Again the stars are undisturbed, while the ionized gas component is, and within the disk they both display projected rotations of $\sim 150$ \kms.
As in A2744 09 we notice the presence of a bump of gas with a velocity consistent with the one shown by the remaining part of the tail, but spatially detached.
Both the gas and the stellar velocity dispersions are quite high, with values reaching $\sim 100$ \kms for the gas, and correspondingly lower for the stars in the galaxy center.
The extinction is generally lower than 1 mag within the stellar disk that is completely star forming. The BPT diagram allows to classify the stripped tail as composite, with some spaxels revealing possible shocks.

\item
A370 02 is a small galaxy with signs of a disturbed morphology, especially towards the east side of the disk. The red galaxy covering large part of the cubelet is not interacting with A370 02, as MUSE spectra confirm.
The \Ha kinematics show a rotation of $\sim 60$ \kms and following our definition of galaxy/tails, shown as black ellipse in Fig.\ref{fig:a370_01_02}, a portion of it is already part of what we define a gas tail.
The central gas velocity dispersion is low, while the beam smearing does not allow to clearly disentangle the stellar velocity/velocity dispersion which turn out to have odd values.
The dust extinction is everywhere low, and the galaxy disk is dominated by the SF. Possible signs of a biconical ionization cone are traced by the AGN/Composite/Shocks spaxels that characterize the initial part of the tail and the symmetric region extending toward west.

\item 
A370 03 is another clear example of a ram-pressure stripped galaxy: its stellar kinematics appears undisturbed, with a projected $v_{rot} \sim 60$ \kms and a similar gaseous kinematics within the galaxy extent (black contours in Fig.\ref{fig:a370_03_06}. However, a very long tail of ionized gas appears toward the south-west direction (both in \Ha and in [OII]).  
The gas velocity increases along the tail and reaches very high values (up to 600 \kms). No other neighbouring galaxies are visible in the HST images.
The tail is characterized by high gas velocity dispersion and a combination of composite and star-forming regions, while the disk is all star-forming According to the BPT.

\item 
A370 06 is clearly a rotation-dominated (with projected velocity of $\sim 150$ \kms) late type galaxy, with a  $\sim 20$ kpc ionized gas tail 
extending toward the south east direction. Just outside the stellar contour traced by the  ellipse another object is seen in projection, locate at a higher redshift, which is not responsible for the tail emission, nor is interacting with A370 06.
The gas in the tail has a very high velocity ($\sim 600$ \kms), and
as can be seen in the  lowest right panel of Fig. \ref{fig:a370_03_06}, it is dominated by AGN-like line ratios, indicating the possible contribution by a central AGN in the development of this tail (possibly an outflow from the AGN).

\item 
A370 07 is a fuzzy disturbed galaxy with a hint of a tail in the north-west direction (see Fig.\ref{fig:RGB}).
The detection is mostly tentative, but its characteristics (i.e. the [OII]/\Ha ratio, see Sec.\ref{sec:o2ha}) is related to an [OII] rich tail left behind from a stripped galaxy.

\item 
A370 08 is the galaxy with the longest tail of ionized gas ($\sim 12$arcsec, i.e. $\sim 57$ kpc).
We carefully checked all the objects located along the tail and especially in the southmost part of the MUSE cubelet, but found no  redshift correspondence with the ionized gas emission which shows a great level of continuity moving from the inner part of the galaxy to the extremes of the tail. The galaxy itself shows a projected rotation of $\sim 200$ \kms, while the gas in the tail reaches $\sim 600$ \kms. 
The north-west side of the galaxy disk is already gas depleted, as is typical if the gas is removed due to ram-pressure stripping.
The gas velocity dispersion is chaotic within the disk, with lower values at the compression front, and higher values at the opposite side and along the tail.
The extinction is generally low, and the region is mostly composite/star forming according to the BPT diagram.

\item 
A370 09 also shows a very faint tail of gas elongated in the north-east to south-west direction (see also Fig. \ref{fig:rps_o2}). 
The analysis of the MUSE cubelet shows hint for such stripped gas emission, and again (as in the case of A370 07) its peculiar characteristics in terms of [OII]/\Ha line ratio, suggesting a ram-pressure origin.

\end{itemize}

\end{document}